\documentclass[12pt]{article}

\usepackage[utf8]{inputenc}
\usepackage{textcomp}
\usepackage{mathtools}
\usepackage{amsmath}
   \usepackage{amssymb,bm}
\usepackage[shortlabels]{enumitem}
\usepackage{chetnew}
\usepackage{tikz}
\usepackage{amssymb,amsfonts,mathrsfs,dsfont,yfonts,bbm}\usetikzlibrary{calc}
\usepackage{xcolor}
\usepackage{braket}
\usepackage[normalem]{ulem}

\newcommand{\beqn}{\begin{eqnarray}}
\newcommand{\eeqn}{\end{eqnarray}}

\newcommand{\be}{\begin{equation}}
\newcommand{\ee}{\end{equation}}
\newcommand{\bea}{\begin{eqnarray}}
\newcommand{\eea}{\end{eqnarray}}

\newcommand{\CH}{\mathcal{H}}

\newcommand{\CD}{\mathcal{D}}

\newcommand{\CQ}{\mathcal{Q}}
\newcommand{\CB}{\mathcal{B}}
\newcommand{\CC}{\mathcal{C}}

\newcommand{\CO}{\mathcal{O}}
\newcommand{\CT}{\mathcal{T}}
\newcommand{\CI}{\mathcal{I}}
\newcommand{\CN}{\mathcal{N}}
\newcommand{\CS}{\mathcal{S}}
\newcommand{\CM}{\mathcal{M}}

\newcommand*{\boxcoloro}{orange}
\makeatletter
\newcommand{\boxedo}[1]{\textcolor{\boxcoloro}{%
\tikz[baseline={([yshift=-1ex]current bounding box.center)}] \node [rectangle, minimum width=1ex,rounded corners,draw] {\normalcolor\m@th$\displaystyle#1$};}}
 \makeatother
\newcommand*{\boxcolorr}{red}
\makeatletter
\newcommand{\boxedr}[1]{\textcolor{\boxcolorr}{%
\tikz[baseline={([yshift=-1ex]current bounding box.center)}] \node [rectangle, minimum width=1ex,rounded corners,draw] {\normalcolor\m@th$\displaystyle#1$};}}
 \makeatother
\newcommand*{\boxcolorb}{blue}
\makeatletter
\newcommand{\boxedb}[1]{\textcolor{\boxcolorb}{%
\tikz[baseline={([yshift=-1ex]current bounding box.center)}] \node [rectangle, minimum width=1ex,rounded corners,draw] {\normalcolor\m@th$\displaystyle#1$};}}
 \makeatother
\newcommand*{\boxcolorg}{green}
\makeatletter
\newcommand{\boxedg}[1]{\textcolor{\boxcolorg}{%
\tikz[baseline={([yshift=-1ex]current bounding box.center)}] \node [rectangle, minimum width=1ex,rounded corners,draw] {\normalcolor\m@th$\displaystyle#1$};}}
 \makeatother
 \newcommand*{\boxcolorp}{purple}
\makeatletter
\newcommand{\boxedp}[1]{\textcolor{\boxcolorp}{%
\tikz[baseline={([yshift=-1ex]current bounding box.center)}] \node [rectangle, minimum width=1ex,rounded corners,draw] {\normalcolor\m@th$\displaystyle#1$};}}
 \makeatother
  \newcommand*{\boxcolorc}{cyan}
\makeatletter
\newcommand{\boxedc}[1]{\textcolor{\boxcolorc}{%
\tikz[baseline={([yshift=-1ex]current bounding box.center)}] \node [rectangle, minimum width=1ex,rounded corners,draw] {\normalcolor\m@th$\displaystyle#1$};}}
 \makeatother
  \newcommand*{\boxcolory}{yellow}
\makeatletter
\newcommand{\boxedy}[1]{\textcolor{\boxcolory}{%
\tikz[baseline={([yshift=-1ex]current bounding box.center)}] \node [rectangle, minimum width=1ex,rounded corners,draw] {\normalcolor\m@th$\displaystyle#1$};}}
 \makeatother

\usepackage{amsmath}	
\preprint{QMUL-PH-21-57, NITEP 126, OCU-PHYS 553\\ \vspace*{-1.5cm}}

\title{Spin Thresholds, RG Flows, and \\[3mm]  Minimality in 4D $\CN=2$ QFT}

\author{Matthew Buican$^{\diamondsuit,1}$, Hongliang Jiang$^{\clubsuit,1}$, and Takahiro Nishinaka$^{\heartsuit,2}$}
\affiliation{\smallskip $^{1}$CTP and Department of Physics and Astronomy\\
Queen Mary University of London, London E1 4NS, UK\\ $^{2}$Department of
Physics/NITEP\\ Osaka City University, Osaka 558-8585, Japan  \emails{$^{\diamondsuit}$m.buican@qmul.ac.uk, $^{\clubsuit}$h.jiang@qmul.ac.uk, $^{\heartsuit}$nishinaka@osaka-cu.ac.jp}}

\abstract{Long ago, Argyres and Douglas discovered a particularly simple interacting 4D $\CN=2$ superconformal field theory (SCFT) on the Coulomb branch of $SU(3)$ $\CN=2$ super Yang-Mills. Further hints of the theory's simplicity arise due to the fact that it has the smallest possible value of the $c$ central charge among unitary interacting $\CN=2$ SCFTs. A main purpose of this note is to uncover additional aspects of this minimal Argyres-Douglas (MAD) theory's simplicity. In particular, we argue that: (1) the MAD theory shares an infinite set of large spin thresholds in part of its operator spectrum with the free $\CN=2$ Maxwell theory (this data is therefore invariant under generic $\CN=2$-preserving renormalization group flows to the IR) and (2) the MAD theory has, at every order in the natural grading, the smallest number of \lq\lq Schur" operators of any unitary $\CN=2$ theory (interacting or free). We then show that property (1) has a suitable generalization for all $(A_1, A_{2k})$ cousins of the MAD theory. In particular, the corresponding large spin thresholds encode generic renormalization group flows within this class. This construction therefore gives a different handle on these flows from the one provided by the Seiberg-Witten description. To emphasize the importance of these spin thresholds, we abstractly study theories with \lq\lq enough matter" to form Higgs branches and argue that infinitely many spin thresholds are small or vanishing.}
\begin{document}

\maketitle
\toc

\newsec{Introduction}
One lesson of the last decade of research into strongly coupled 4D $\CN=2$ SCFTs is that these theories often have a hidden simplicity. An important tool for revealing this simplicity is the mapping between 4D $\CN=2$ SCFTs and 2D chiral algebras \cite{Beem:2013sza}. The corresponding 2D theories are often tightly constrained: they typically have a simple set of generators (e.g., see \cite{Beem:2013sza,Lemos:2014lua,Buican:2015ina,Cordova:2015nma,Buican:2016arp,Xie:2016evu,Creutzig:2017qyf,Song:2017oew,Buican:2017fiq,Buican:2017rya,Creutzig:2018lbc,Bonetti:2018fqz,Arakawa:2018egx,Xie:2019yds,Xie:2019zlb,Beem:2019snk,Xie:2019vzr}) and obey interesting modular relations \cite{Beem:2017ooy}.

Another aspect of the simplicity that reveals itself through the map in \cite{Beem:2013sza} is that strongly interacting 4D theories often have surprising relations with massless free fields \cite{Buican:2017rya,Bonetti:2018fqz,Beem:2019tfp,Pan:2021ulr}. While the full physical implications of these results are not understood, it is clear that there are simplifying principles in 4D $\CN=2$ SCFTs yet to be identified.

One purpose of this note is to further explore massless free field relations with interacting SCFTs.\footnote{Unless otherwise stated, any free fields we discuss below are massless.} In particular, the above relations, which proceed through the so-called Schur sector of 4D operators subject to the map in \cite{Beem:2013sza}, involve Higgs branches either explicitly or implicitly.\footnote{By \lq\lq Higgs branch" we mean the potentially more general set of vacua where the UV superconformal $SU(2)_R$ is spontaneously broken. In other words, this can include branches of vacua where free vector multiplets or interacting IR components appear in addition to the axion-dilaton hypermultiplet.} However, not all interacting 4D theories have Higgs branches. On the other hand, all such theories are believed to have Coulomb branches. It is therefore important to find relations between free fields and interacting theories that only possess a Coulomb branch.

If one ignores the Schur sector that features in the 4D/2D relation of \cite{Beem:2013sza}, this is, in some sense, what has been done through the research program of reconstructing SCFTs from their (generically) IR-free Coulomb branches. Indeed, great progress has been made in this direction starting long ago with \cite{Seiberg:1994aj} and various generalizations (e.g., see \cite{Argyres:2015ffa,Argyres:2015gha,Moore:2017cmm,Martone:2020nsy,Argyres:2020wmq,Cecotti:2021ouq} for a small subset of interesting recent results). This route tends to primarily constrain the 4D $\CN=2$ chiral sector of the UV theory (i.e., the set of UV operators annihilated by all the anti-chiral $\CN=2$ supercharges) and mostly ignore the Schur sector.\footnote{However, this program also makes contact with flavor symmetry and the associated $\CN=2$ (generalized) mass deformations; it can also capture other aspects of Higgs branch physics.}

Clearly, if our goal is to fully understand $\CN=2$ SCFTs, we must understand how the chiral and Schur sectors interact. In that vein, one more modest goal of the present paper is to relate (massless) Coulomb branch physics with the Schur sector in certain 4D $\CN=2$ SCFTs.

While this goal may seem slightly quixotic, it is certainly not without precedent. For example, the $S^1$ reductions of the Schur indices of various 4D $\CN=2$ SCFTs know about the $U(1)_r$ quantization of the $\CN=2$ chiral operators \cite{Buican:2015hsa,Fredrickson:2017yka,Dedushenko:2019mnd}. Importantly, the Schur index formula of \cite{Cordova:2015nma} proceeds via a counting of (generically massive) Coulomb branch BPS states (see also closely related constructions in \cite{Gaiotto:2010be,Cecotti:2010fi,Iqbal:2012xm,Cecotti:2015lab}).\footnote{The main difference in our approach will be to construct operators (and null states) in the massless IR theory on the Coulomb branch (including cases where the IR is interacting) with the same quantum numbers as ones in the UV.} Finally, certain 3D TQFTs discussed in \cite{Dedushenko:2018bpp} seem to combine data from the Schur and chiral sectors of a theory (see also \cite{Buican:2019huq}).

A natural starting point for such an exploration is the original Argyres-Douglas (AD) theory discovered in \cite{Argyres:1995jj}. Indeed, it lacks a Higgs branch, and, from the effective IR Coulomb branch perspective, it is the simplest possible interacting SCFT: it can be roughly understood as a point on the Coulomb branch of a non-abelian gauge theory where there is a single massless abelian vector multiplet coupled to two massless hypermultiplets with mutually non-local electric and magnetic charges. Since abelian gauge theories with purely electrically charged matter are IR-free in 4D, this picture heuristically suggests that the AD theory of \cite{Argyres:1995jj} is the simplest interacting $\CN=2$ SCFT. We will therefore refer to it as the \lq\lq minimal AD theory ", or \lq\lq MAD theory" for short. As a note to readers, we have in more common terminology
\begin{equation}\label{equivth}
{\rm MAD\ SCFT} = (A_1, A_2)\ {\rm SCFT} = H_0\ {\rm SCFT}={\rm AD}_{N_f=0}(SU(3))={\rm AD}_{N_f=1}(SU(2))~. 
\end{equation}

Intriguingly, an analysis of the Schur sector of the MAD theory shows it has the smallest possible value of $c$ for any unitary interacting 4D $\CN=2$ SCFT \cite{Liendo:2015ofa}. Moreover, the MAD Schur sector turns out to be isomorphic to the chiral algebra of the Lee-Yang theory \cite{Cordova:2015nma}; this latter theory is the simplest Virasoro minimal model.\footnote{We will make this notion more precise in what follows. One way to see this statement physically is to note that the Lee-Yang theory can be reached via a (unitarity-violating) RG flow from the 2D Ising CFT. More generally, it has the smallest value of $c_{\rm eff}$ among all Virasoro minimal models and so the result in \cite{Castro-Alvaredo:2017udm} implies it is \lq\lq simplest."} Therefore, from this perspective, the MAD SCFT is a particularly simple theory.\footnote{There may be other SCFTs with the same $c$. However, since the 1-form symmetry of the $(A_1, A_2)$ theory is trivial \cite{Closset:2020scj,DelZotto:2020esg,Closset:2020afy,Closset:2021lwy}, we do not expect the situation we have in, say, $\CN=4$ SYM where we have theories with the same local structure and different 1-form symmetry (e.g., see the discussion in \cite{Buican:2021xhs}). Still, there could be theories with the same $c$ and more subtle global differences (or even different sets of local observables).}

Our first result, presented in section \ref{A1A2spin}, unifies this Schur sector perspective on the MAD theory's simplicity with the one arising from the Coulomb branch effective action. More precisely, we argue that an infinite set of spin thresholds that appear in the MAD Schur sector for each value of the $SU(2)_R$ weight are precisely reproduced by the free vector multiplet---the massless theory on the Coulomb branch.\footnote{As an aside, we note that this information naively goes beyond the data directly associated with the chiral algebra, since $SU(2)_R$ charge is not respected in the 4D/2D relation of \cite{Beem:2013sza}. However, see \cite{Song:2016yfd,Beem:2019tfp} for some highly non-trivial results in reconstructing $SU(2)_R$ from the 2D perspective.} Therefore, the MAD spin threshold data is invariant under turning on vevs for $\CN=2$ chiral ring operators.

If we also turn on an $\CN=2$-preserving relevant deformation, then the Coulomb branch is deformed. At special co-dimension one points on this moduli space, the massless IR theory is $N_f=1$ $\CN=2$ SQED (the \lq\lq$I_1$" theory in the nomenclature of \cite{Argyres:2015ffa}). In this case, the IR theory has vanishing spin thresholds, and the UV/IR spin threshold equality becomes an inequality. More generally, we expect infinitely many IR thresholds to be less than or equal to UV thresholds for theories in which there is an irrelevantly gauged flavor symmetry in the IR (e.g., as in SQED or as in the \lq\lq quantum Higgs branch" examples of \cite{Argyres:2012fu}).\footnote{By irrelevantly gauged flavor symmetry, we mean that the corresponding gauge coupling is IR free.} It would be interesting to understand if these  are the only such cases.\footnote{Through out this work, when we study Coulomb branches we have in mind moduli spaces with only free vectors at generic points (i.e., not so-called \lq\lq enhanced" Coulomb branches).}

After establishing the above picture, we then generalize our discussion to the infinite set of $(A_1, A_{2k})$ theories, of which the MAD theory is the simplest (i.e., $k=1$). In particular, we show that the corresponding spin thresholds are invariant under generic $\CN=2$-preserving deformations that take us within this class of theories (again, as long as there are no irrelevant gaugings appearing in the IR). 

In section \ref{genspin}, we consider how the picture changes in theories that support Higgs branches. In particular, we give some universal constraints on the Schur sector of such theories that follow from locality and argue that, unlike the case of $(A_1, A_{2k})$ theories, infinitely many of the corresponding Schur sector spin thresholds are small or vanish.

The second main point of this paper is that the MAD SCFT has the simplest possible Schur sector for any unitary 4D SCFT. This result, presented in section \ref{minSchur}, is a generalization of a result in \cite{Buican:2016arp}. There, one sees that the MAD SCFT has the smallest asymptotic Schur index for any 4D $\CN=2$ SCFT with purely bosonic Schur operators. One upshot of \cite{Buican:2016arp} was that the MAD theory has, in addition to the smallest $c$ for a unitary interacting $\CN=2$ SCFT, the smallest positive $c-a$ for any local $\CN=2$ SCFT.\footnote{Of course, there are theories with $c-a=0$ as in the case of $\CN=4$ SYM (and there are also theories with $c-a<0$).}

Going beyond this result, we show in section \ref{minSchur} that, in fact, the MAD SCFT has the smallest Schur index---at each order in the grading---of any unitary 4D $\CN=2$ SCFT with a purely bosonic Schur sector. One corollary of this proof is the result that, for any unitary 4D $\CN=2$ SCFT, the MAD theory has the smallest number of Schur operators at each order in the grading (where we sum the number of bosonic and fermionic operators at a given order without weighting by fermion number).

Before getting to these results we give a brief review of the Schur sector and the chiral algebra construction of \cite{Beem:2013sza} in the next section. After presenting the above results in sections \ref{A1A2spin}, \ref{genspin}, and \ref{minSchur}, we conclude with a discussion of open problems and a conjecture on the maximality of the spin thresholds present in the MAD SCFT.

\newsec{A brief review of the Schur sector}\label{review}
In this section we briefly review the Schur sector and the related construction of \cite{Beem:2013sza}. We do this with a view toward emphasizing the aspects that will be particularly important to us below.

To that end, we first note that Schur operators live in certain short multiplets of $\CN=2$ superconformal symmetry and satisfy
\begin{equation}\label{Schurcond0}
\left\{\CQ_-^1,\CO\right]=\left\{\tilde\CQ_{2\dot-},\CO\right]=\left\{\CS^-_1,\CO\right]=\left\{\tilde\CS^{2\dot-},\CO\right]=0~,
\end{equation}
where the numerical labels indicate $SU(2)_R$ weight (a raised \lq\lq1" indicates highest weight of the spin-half representation), and \lq\lq$-,\dot-$" are weights of Euclidean spin. Here the $\CQ$'s are Poincar\'e supercharges, and the $\CS$'s are special supercharges. It turns out that \eqref{Schurcond0} is enough to guarantee that
\begin{equation}\label{Schurcond}
\Delta(\CO)=2R(\CO)+j(\CO)+\bar j(\CO)~,\ \ \ r(\CO)=\bar j(\CO)-j(\CO)~.
\end{equation}
In \eqref{Schurcond}, $\Delta$ is the scaling dimension, $R$ is the $SU(2)_R$ weight, $r$ is the $U(1)_r$ charge, and $j,\bar j$ denote the left and right spin weights. These quantum numbers are precisely right for the operators to contribute to the Schur limit of the superconformal index \cite{Gadde:2011uv}\footnote{We drop possible refinements by flavor fugacities, as the MAD theory and its close cousins lack flavor symmetries.}
\begin{equation}
\CI_S^{\CT}(q):={\rm Tr}(-1)^Fq^{\Delta-R}~,
\end{equation}
where the trace is over operators satisfying \eqref{Schurcond}, $(-1)^F$ is fermion number, and $\CT$ denotes the SCFT in question.

It turns out that all Schur operators live in one of the following four types of multiplets\footnote{We use the notation of \cite{Dolan:2002zh}; see also the discussion in \cite{Dobrev:1985qv,Cordova:2016emh}.}
\begin{equation}
\hat\CB_R~,\ \ \ \CD_{R(0,\bar j)}~,\ \ \ \bar\CD_{R(j,0)}~,\ \ \ \hat\CC_{R(j,\bar j)}~.
\end{equation}
The $\hat\CB_R$ multiplets include the flavor symmetry currents when $R=1$ (the Schur operator is the holomorphic moment map) and all Higgs branch operators, while the $\CD_{R(0,\bar j)}\oplus\bar\CD_{R(j,0)}$ multiplets include free vectors and extra supercurrents among others. These three multiplets together comprise the so-called Hall-Littlewood (HL) subsector of the Schur sector (we can think of $\CD_{R(0,\bar j)}$ as contributing to the HL anti-chiral ring and $\bar\CD_{R(j,0)}$ as contributing to the HL chiral ring, while $\hat\CB_R$ contributes to both).

On the other hand, the $\hat\CC_{R(j,\bar j)}$ multiplets are, in a sense we will see below, the most \lq\lq generic" or \lq\lq universal" types of Schur multiplets. The most famous among them is the $R=j=\bar j=0$ stress-tensor multiplet, which is present in any local theory. The Schur operator is the level-two descendant corresponding to the highest-$SU(2)_R$ and Euclidean spin weight component of the $SU(2)_R$ current, $J^{11}_{+\dot+}$. Note that, more generally, the $\hat\CC_{R(j,\bar j)}$ primary has quantum numbers $(R,j,\bar j)$, while the associated Schur operator is a level-two superconformal descendant and transforms as the highest-weight component of the representation $(R+1,j+1/2,\bar j+1/2)$.

The other $\hat\CC_{R(j,\bar j)}$ multiplets (with the exception of the $\hat\CC_{0(j,\bar j)}$ higher-spin current multiplets in free theories\footnote{Note that here all the caveats of \cite{Maldacena:2011jn,Alba:2015upa} apply: in particular, the correlation functions of the stress tensor and conserved currents are those of a free theory even if there is no free field in the spectrum.}) are often forgotten when thinking about 4D $\CN=2$ SCFTs. One of the main points of this paper is that {\it spin thresholds in the $U(1)_r$-neutral $\hat\CC_{R(j,j)}$ multiplet sector shed interesting light on RG dynamics of strongly interacting theories.}

\subsection{The 4D/2D correspondence}
Given the above description of the Schur sector, we can now introduce the 4D/2D correspondence of \cite{Beem:2013sza}. This construction starts from the observation that \eqref{Schurcond0} is equivalent to the statement that
\begin{equation}\label{Schurcond2}
\left\{\mathbbmtt{Q}_i,\mathcal{O}(0)\right]=0~, \ \ \ \mathcal{O}(0)\ne\left\{\mathbbmtt{Q}_i,\mathcal{O}'(0)\right]~,\ \ \ \mathbbmtt{Q}_1:=\CQ^1_-+\tilde S^{2\dot-}~, \ \ \ \mathbbmtt{Q}_2:=S_1^{-}-\tilde Q_{2\dot-}~.
\end{equation}
In this light, the Schur conditions become a statement that Schur operators are representatives of non-trivial cohomology classes with respect to the $\mathbbmtt{Q}_i$ (note that these charges satisfy $\mathbbmtt{Q}_i^2=0$). From now on, we will drop the subscript and simply write $\mathbbmtt{Q}_i\to\mathbbmtt{Q}$ since the cohomology does not depend on $i$.

The main idea of \cite{Beem:2013sza} is to then fix the Schur operators to a plane, $\mathcal{P}\subset\mathbb{R}^4$, and twist the right-moving global conformal transformations on $\mathcal{P}$ with $SU(2)_R$ while leaving the left-moving transformations untouched
\begin{eqnarray}\label{2dgens}
L_{-1}&=&\mathcal{P}_{+\dot+}~,\ \ \ L_1=\mathcal{K}^{+\dot+}~,\ \ \ L_0={1\over2}(\mathcal{H}+\CM_+^{\ +}+\CM^{\dot+}_{\ \dot+})~,\cr \widehat{L}_{-1}&=&\mathcal{P}_{-\dot-}+\mathcal{R}^-~,\ \ \ \widehat{L}_1=\mathcal{K}^{-\dot-}-\mathcal{R}^+~,\ \ \ \widehat L_0={1\over2}(\mathcal{H}-\CM_+^{\ +}-\CM^{\dot+}_{\ \dot+})-\mathcal{R}~.
\end{eqnarray}
Here $\mathcal{P}_{\alpha\dot\alpha}$ is the generator of translations, $\mathcal{K}_{\alpha\dot\alpha}$ is the generator of special conformal transformations, $\CH$ is dilation, $\mathcal{R}, \mathcal{R}^{\pm}$ are the $SU(2)_R$ generators, and $\CM_{\alpha}^{\ \beta},\CM^{\dot\beta}_{\dot\alpha}$ generate rotations.

The key point is that the $\widehat{L}_i$ are in fact $\mathbbmtt{Q}$-exact (and therefore also $\mathbbmtt{Q}$-closed, i.e., they commute with $\mathbbmtt{Q}$). As a result, translating the Schur operators in $\mathcal{P}$ using the generators in \eqref{2dgens} does not change the $\mathbbmtt{Q}$ cohomology class and, moreover, if we work in $\mathbbmtt{Q}$ cohomology, the coordinate dependence of the twisted-translated operators is holomorphic. This is the hallmark of a 2D chiral algebra.

Some examples of this map include
\begin{equation}
\chi([J^{11}_{+\dot+}]_{\mathbbmtt{Q}})=T~,\ \ \ \chi([\mu]_{\mathbbmtt{Q}})=J~, \ \ \ \chi(\partial_{+\dot+})=\partial_z:=\partial~,
\end{equation}
where $J^{11}_{+\dot+}$ is the Schur operator in the stress tensor multiplet, $T$ is the 2d holomorphic stress tensor, $\mu$ is the holomorphic moment map for some flavory symmetry, $J$ is the related 2D affine current, and \lq\lq$[\cdots]_{\mathbbmtt{Q}}$" denotes the $\mathbbmtt{Q}$-cohomology class of the enclosed operator.

Given this discussion, it is then natural that the torus partition sum of the 2D chiral algebra, $Z$, gives the Schur index
\begin{equation}\label{equivcount}
\CI_S(q)=Z(-1,q)~,\ \ \ Z(y,q):={\rm Tr}y^{M^{\perp}}q^{L_0}~, 
\end{equation}
where $y$ is a fugacity for rotations normal to the chiral algebra plane (this is a non-locally realized symmetry of the 2D theory). Here the $L_0$ eigenvalue is commonly denoted as $h=\Delta-R$. Another useful point to make about the equivalent 4D and 2D counting in \eqref{equivcount} is that conformal primary Schur operators in 4D are mapped to $sl(2,\mathbb{R})$ primaries in 2D (this follows from the second equation in \eqref{2dgens}). Conversely, if the 4D Schur operator is a descendant, then it can be written as $\CO=\partial_{+\dot+}^n\CO'$, where $\CO'$ is a primary Schur operator. Working at the origin of $\mathcal{P}$, we see that $[\CO]_{\mathbbmtt{Q}}=\partial^n[\CO']_{\mathbbmtt{Q}}$, and so the corresponding chiral algebra state is an $sl(2,\mathbb{R})$ descendant.

Since we have twisted with $SU(2)_R$, the chiral algebra naively looses information about the $SU(2)_R$ weight of the 4D Schur operator.\footnote{In some cases, a prescription is known for how to recover this information \cite{Song:2016yfd,Beem:2019tfp}, but we will not need to use these methods below.} The main manifestation of this fact for us is that 4D OPEs of Schur operators are mapped in cohomology to \cite{Beem:2013sza}
\begin{equation}
[\CO_1]_{\mathbbmtt{Q}}(z)[\CO_2]_{\mathbbmtt{Q}}(0)=\sum_{k\in{\rm Schur}}{\lambda_{12k}\over z^{h_1 +h_2-h_k}}[\CO_k]_{\mathbbmtt{Q}}(0)~.
\end{equation}
In particular, Schur operators with the same $h=\Delta-R$ but different $R$ can contribute at the same order in the 2D OPE (e.g., the twisting therefore changes the definition of 2D normal ordering relative to 4D). This is an ambiguity that requires care to resolve when making statements about 4D. 

Finally, let us conclude by noting that a consequence of this correspondence is that
\begin{equation}\label{eq:centralcharges}
c_{2d}=-12c~,\ \ \ k_{2d}=-{1\over2}k~,
\end{equation}
where $c_{2d}$ and $k_{2d}$ are the 2D central charge and affine level respectively, and $c,k$ are the corresponding 4D quantities. As a result, unitary 4D theories map to non-unitary 2D theories and vice-versa.

\subsec{The free abelian vector multiplet}
Since a main point of our discussion below will be to compare the free $\CN=2$ Maxwell theory's Schur sector with the MAD Schur sector, we briefly review how to construct Schur operators for the free abelian vector multiplet.

To that end, all Schur operators are generated by $\lambda^1_+$ and $\bar\lambda^1_{\dot+}$. In particular, we have
\begin{equation}\label{genMaxSchur}
\CO_{n,\bar n, j, \bar j}:=\partial_{+\dot+}^{i_1}\lambda_+^1\partial^{i_2}_{+\dot+}\lambda_+^1\cdots\partial_{+\dot+}^{i_n}\lambda_+^1\cdot\partial_{+\dot+}^{k_1}\bar\lambda_{\dot+}^1\partial^{k_2}_{+\dot+}\bar\lambda_{\dot+}^1\cdots\partial_{+\dot+}^{k_{\bar n}}\bar\lambda_{\dot+}^1~.
\end{equation}
Since the gauginos transform as follows under $(SU(2)_R,U(1)_r,SU(2)_j,SU(2)_{\bar j})$
\begin{equation}
\lambda_+^1\oplus\bar\lambda^1_{\dot+}\in(1/2,-1/2,1/2,0)\oplus(1/2,1/2,0,1/2)~,
\end{equation}
the operator in \eqref{genMaxSchur} transforms as
\begin{equation}
\CO_{n,\bar n, j, \bar j}\in\left({n+\bar n\over2},{-n+\bar n\over2},j,\bar j\right)~,
\end{equation}
with
\begin{equation}
j={1\over2}\left(n+\sum_{a=1}^ni_a+\sum_{b=1}^{\bar n}k_b\right)~,\ \ \ \bar j={1\over2}\left(\bar n+\sum_{a=1}^ni_a+\sum_{b=1}^{\bar n}k_b\right)~.
\end{equation}

Note that, since all Schur operators in this theory have spin, there are no $\hat\CB_R$ multiplets. This fact is to be expected: the free vector multiplet has no Higgs branch. However, it is possible to produce conformal primaries
\begin{equation}
\CO_{1,0,1,0}\in\bar\CD_{0(0,0)}~,\ \ \ \CO_{0,1,0,1}\in\CD_{0(0,0)}~,\ \ \ \CO_{n,\bar n,j,\bar j}\in\hat\CC_{(n+\bar n)/2-1(j-1/2,\bar j-1/2)}~, \ \ \ n+\bar n\ge2~.
\end{equation}
In particular, we learn that the vast majority of free vector Schur operators are of type $\hat\CC_{R(j,\bar j)}$. This fact is consistent with the statement that the $\hat\CC_{R(j,\bar j)}$ operators are the most generic Schur operators.\footnote{Although it will not play much of a role in our discussion below, we note in passing that the $c_{2d}=-2$ chiral algebra associated with the free $\CN=2$ Maxwell theory is the small algebra of the $(b,c)$ system of weight $(1,0)$ \cite{Beem:2013sza}.} 

A particularly important part of our story below will be played by the $U(1)_r$-neutral Schur operators. In the case of the free vector multiplet these are the primaries of type
\begin{equation}
\CO_{n,n,j,j}\in\hat\CC_{(n+\bar n)/2-1,(j-1/2,j-1/2)}~.
\end{equation}
As we will see, this (infinite) subsector will carry imprints of RG flows from the MAD theory and its higher-rank cousins. 

\newsec{Spin thresholds in the MAD Schur sector and the free $\CN=2$ super Maxwell theory}\label{A1A2spin}
In this section we wish to compare the 4D Schur sector of the MAD theory with the Schur sector of the free $\CN=2$ vector multiplet. In particular, we will argue that an infinite amount of 4D data in these two sectors agrees.\footnote{That there is a relation between these Schur sectors is not entirely unexpected. Indeed, \cite{Cordova:2015nma} constructs the MAD Schur index via contributions from the free vector multiplet dressed with various massive BPS contributions. Here we focus on operators in the UV and massless IR theories and find surprisingly direct relations between infinite sets of 4D quantum numbers.}

As mentioned in the introduction, the (minimal) effective Coulomb branch description of the MAD theory includes a massless abelian vector multiplet coupled to two massless hypermultiplets with mutually non-local electric/magnetic charges. The resulting theory has no flavor symmetry and no Higgs branch. Moreover, this description shows that there is no standard $\CN=2$ Lagrangian for the MAD SCFT. This point is further driven home by the fact that the generator of the MAD chiral ring is an operator, $\CO_{6/5}$, with scaling dimension $6/5$. Since this scaling dimension is non-integer, the theory must be non-Lagrangian (in a theory with an $\CN=2$ Lagrangian such operators correspond to Casimirs built out of the vector multiplet scalars).\footnote{Intriguingly, $6/5$ is not too far off from the free field scaling dimension of $1$. Indeed, this fact may be related to the free field imprints we find below.} Therefore, constructing the Schur sector in this case cannot be as easy as it was in the case of the free vector multiplet. Instead, our understanding of this sector arises from various more indirect pieces of evidence we summarize below.

For our purposes, it is useful to present the data of the 4D MAD Schur sector in a way that is simpler than what has appeared in the literature to date. In particular, we begin by arguing for the following claim:

\medskip\noindent
{\bf Claim 1:} The Schur spectrum of the MAD theory consists exclusively of $\hat\CC_{R(j,j)}$ multiplets with generating function (for $R>0$)
\begin{equation}\label{ChatspecMAD}
f_{\hat\CC_{R(j,j)}}(q)={q^{R(R+2)}\over(1-q^2)(1-q^3)\cdots(1-q^{R+1})}=\sum_{j=0}^{\infty}N_{\hat\CC_{R(j,j)}}q^{2j}~, \ \ \ R\in\mathbb{Z}_{>0}~,
\end{equation}
where $N_{\hat\CC_{R(j,j)}}$ is the number of $\hat\CC_{R(j,j)}$ Schur multiplets. For $R=0$, we set $f_{\hat\CC_{0(j,j)}}=N_{\hat\CC_{0(j,j)}}=\delta_{0j}$ since there is a unique stress tensor multiplet, and the theory is interacting (it therefore doesn't contain higher-spin conserved currents). Note that there are no MAD Schur multiplets with half-integer $SU(2)_R$ spin primaries. Moreover, since the left and right spins of the primaries are equal, all Schur operators are $U(1)_ r$-neutral bosons.

\medskip
To support this claim, let us first recall that, as alluded to in the introduction, the Schur sector of the MAD theory (really the corresponding cohomology as reviewed in section \ref{review}) is isomorphic to the Lee-Yang chiral algebra (i.e., the Virasoro algebra at $c_{2d}=-22/5$). This statement is strongly suggested by the Schur index construction in  \cite{Cordova:2015nma}.

While this result doesn't immediately tell us which 4D operator a given Lee-Yang operator corresponds to, it severely constrains the possibilities. In particular, we cannot have any $\hat\CB_R$, $\CD_{R(0,j_2)}$, or $\bar\CD_{R(j_1,0)}$ Schur operators in this case. The reason is that such operators correspond to non-trivial Virasoro primaries \cite{Beem:2013sza} which are by definition absent in the Lee-Yang chiral algebra. Moreover, we cannot have any operators $\hat\CC_{R(j_1,j_2)}$ with $j_1\ne j_2$. The reason is that then the corresponding Schur operator will have non-trivial $U(1)_r$ charge (the same logic again rules out $\CD_{R(0,j_2)}\oplus\bar\CD_{R(j_1,0)}$). However, we expect that all Schur operators in this theory arise from the $n$-fold OPE (for $n\ge2$) of the highest Euclidean spin and $SU(2)_R$ weight component of the $SU(2)_R$ current (i.e., the Schur operator of the $\hat\CC_{0(0,0)}$ stress tensor multiplet).\footnote{The two-fold product consists of all Schur operators appearing in the OPE of the $SU(2)_R$ currents. The three-fold product includes the OPE of the $SU(2)_R$ current with all operators appearing in the two-fold product. We continue inductively for all $n>3$. } Since the 4D/2D map preserves $U(1)_r$ (this symmetry is non-locally realized in the chiral algebra), we can only have $U(1)_r$-neutral $\hat\CC_{R(j,j)}$ Schur multiplets in the MAD theory.\footnote{$U(1)_r$-neutrality of the Schur sector also follows from the $\CN=1$ Lagrangian in \cite{Maruyoshi:2016tqk}.}

Now, to derive \eqref{ChatspecMAD}, we may appeal to the Macdonald index. For a general theory, $\CT$, this index counts the same local operators as the Schur index but with an additional fugacity\footnote{In general, there can also be non-trivial flavor fugacities. However, since the MAD theory has no flavor symmetry, we do not bother to include such contributions to the general Macdonald index.}
\begin{equation}
\CI^{\CT}_M(q,T):={\rm Tr}(-1)^Fq^{\Delta-R}T^{R+r}~,
\end{equation}
where the trace is over the space of Schur operators, and we have followed the fugacity conventions of \cite{Agarwal:2018zqi}. In particular, the Schur index corresponds to setting $T=1$.

In the case of the MAD theory, the Macdonald index has been constructed via TQFT in \cite{Song:2015wta} and via the $\CN=1\to\CN=2$ enhancing RG flow in \cite{Maruyoshi:2016tqk}. However, for us, the slightly simpler expression in \cite{Foda:2019guo} will be most useful:
\begin{equation}\label{MADindex}
\CI^{\rm MAD}_M(q,T)=\sum_{n=0}^\infty \frac{q^{n^2+n}}{(q)_n} T^n
 = 1+q^2 T+q^3 T+q^4 T+q^5 T+q^6 \left(T^2+T\right)+q^7 \left(T^2+T\right)+\cdots~,
\end{equation}
where
\begin{equation}\label{qpoch}
(q)_n:=\prod_{i=1}^n(1-q^i)~,\ \ \ (q)_0:=1~.
\end{equation}
To proceed, recall that the Macdonald index for the $\hat\CC_{R(j_1,j_2 )}$ multiplet is\footnote{On the other hand, for the $\hat\CB_R,\CD_{R(0,\bar j)}$,and $\overline{\CD}_{R(j,0)}$ multiplets, the Macdonald index contributions are given by $\CI_{M,\hat\CB_R}(q,T)=  {q^R T^R}/(1-q)$, $\CI_{M,
\CD_{R(0,\bar j)}}(q,T)=  {q^{R+1+\bar j} T^{R+1+\bar j}}/(1-q)$, and $\CI_{M,\bar\CD_{R(j,0)}}(q,T)= {q^{R+1+j} T^{R-1-j}}/(1-q)$. By the general arguments above, these multiplets should not contribute  to the index in \eqref{MADindex}. One can see directly, by multiplying both sides of \eqref{MADindex} with $(1-q)$ (after subtracting the identity contribution) and comparing the $sl(2, \mathbb R)$  primary operators, that $\hat\CB_R$ and $\CD_{R(0,\bar j)}$ are absent (here we use the fact that there are no fermionic operators present in the Lee-Yang vacuum character). CPT invariance implies that $\bar\CD_{R(j,0)}$ is also absent.}
 \be
\CI_{M, \hat {\mathcal C}_{R(j_1,j_2)}}=(-1)^{2(j_1+j_2)}\frac{q^{R+2+j_1+j_2} T^{R+1+j_2-j_1}}{1-q}
 \xrightarrow{j_1=j_2=j}
 \frac{q^{R+2+2j} T^{R+1 }}{1-q}~.
 \ee
Then, using the fact that $f_{\hat\CC_{0(j,j)}}=N_{\hat\CC_{0(j,j)}}=\delta_{0j}$ and adding in the contribution to the index from the identity, we have
\beqn
 1+\sum_{R=0}^\infty \sum_{j=0}^\infty N_{\hat\CC_{R(j,j)}}I_{\hat {\mathcal C}_{R( j ,j)}}
 &=& 1+\sum_{R=0}^\infty \sum_{j=0}^\infty N_{\hat\CC_{R(j,j)}} \frac{q^{R+2+2j} T^{R+1 }}{1-q}
 = 1+\sum_{R=0}^\infty \frac{q^{R+2 }T^{R+1 }}{1-q}\sum_{j=0}^\infty N_{\hat\CC_{R(j,j)}}  q^{2j}\nonumber
 \\ &=&1+{q^2T\over(1-q)}+
 \sum_{R=1}^\infty \frac{q^{R+2 }T^{R+1 }}{1-q}
 \frac{ q^{R(R+2)} }{(1-q^2) (1-q^3) \cdots (1-q^{R+1})}\nonumber
  \\ &=&1+{q^2T\over(1-q)}+
 \sum_{R=1}^\infty \frac{q^{(R+2 )(R+1)}T^{R+1 }}{ (q)_{R+1}}
   \\ &=&
 \sum_{n=0}^\infty \frac{q^{n(n+1)}T^{n }}{ (q)_{n}}=\CI^{\rm MAD}_M(q,T)~,\nonumber
 \eeqn
thereby establishing our claim. Note that in the last equality $n=R+1$, and, in going to the second line, we have used \eqref{ChatspecMAD}.

In what follows, the most interesting consequence of Claim 1 is the following:

\medskip
\noindent
{\bf Consequence (spin thresholds):} For a fixed $SU(2)_R$ representation, the smallest spin $\hat\CC_{R(j,j)}$ primary of the MAD theory is unique and has spin
\begin{equation}\label{MADgaps}
j_{\rm min}(R) = \begin{cases}
{R(R+2)\over2}~, &\text{if}\ R\in\mathbb{Z}_{\ge0}\\
\infty~, &\text{if}\ R\in{1\over2}\mathbb{Z}_{\ge0}~,
\end{cases}
\end{equation}
where, in the half-integer $SU(2)_R$ weight case, we use \lq\lq$\infty$" to denote that such multiplets are absent.

\medskip
\noindent
Clearly $j_{\rm min}(0)=0$ since the theory is local. However, for $R>0$, we see that the minimum spin is non-zero. We refer to the set of $j_{\rm min}(R)>0$ as the spin thresholds of the theory (here we ignore any non-Schur operators). After building more intuition regarding the $\hat\CC_{R(j,j)}$ multiplets in section \ref{genspin}, we will see that it is very easy to engineer interacting theories lacking such thresholds and seemingly much harder to engineer ones that have them for all $R>0$.

Now we wish to compare the spin thresholds in \eqref{MADgaps} with those of the free $\CN=2$ Maxwell theory. To that end, recall from section \ref{review} that all the Schur operators in this theory are constructed from non-vanishing words built out of the gauginos $\lambda_+^1$ and $\bar\lambda_{\dot+}^1$ along with the $\partial_{+\dot+}$ derivative. To construct the $\hat\CC_{R(j,j)}$ Schur operators, we must involve $R+1$ $\lambda_+^1$ and $R+1$ $\bar\lambda_{\dot+}^1$ fields (the Schur operator has total $SU(2)_R$ weight $R+1$) and $2j-R$ factors of $\partial_{+\dot+}$. From this fact alone, we see that $j_{\rm min}(R)\ge{R\over2}$ and that half-integer $SU(2)_R$ weight is forbidden (as in the MAD theory).

To get a stronger result, note that, for a given $R$, the unique minimal spin $\hat\CC_{R(j,j)}$ Schur operator takes the form
\begin{equation}
\CO_{R+1,j_{\rm min}(R)+{1\over2}}=\left(\lambda^1_+\partial_{+\dot+}\lambda^1_+\partial^2_{+\dot+}\lambda^1_+\cdots\partial_{+\dot+}^R\lambda^1_+\right)\left(\bar\lambda^1_{\dot+}\partial_{+\dot+}\bar\lambda^1_{\dot+}\partial^2_{+\dot+}\bar\lambda^1_{\dot+}\cdots\partial_{+\dot+}^R\bar\lambda^1_{\dot+}\right)~.
\end{equation}
Clearly, the total left spin of the corresponding superconformal primary is
\begin{equation}\label{SpinFrVect}
j_{\rm min}(R)={1\over2}(R+1+R(R+1))-{\frac12}={R(R+2)\over2}~.
\end{equation}
In other words, as promised, the free abelian vector multiplet has non-zero spin thresholds for all $R>0$ (here we again ignore non-Schur operators as well as Schur operators with non-trivial $U(1)_r$ charge) and has all its spin thresholds equal to those of the MAD theory
\begin{equation}\label{gapEqMAD}
j_{\rm min}^{\rm free\ vec.}(R)=j_{\rm min}^{\rm MAD}(R)~,\ \ \ \forall R\ge0~.
\end{equation}
Moreover, just as in the MAD theory, there is a unique lowest spin $\hat\CC_{R(j_{\rm min}(R),j_{\rm min}(R))}$ multiplet for every integer $R$.\footnote{Note that the multiplicity of $\hat\CC_{R(j,j)}$ multiplets for fixed $R$ and $j> j_{\rm min}(R)$ is larger in the free vector multiplet case. In fact, this difference is already hinted at by looking at the $R=0$ sector. There, we have a unique multiplet in the case of the MAD theory (the stress tensor multiplet), but we have an infinite set of higher-spin multiplets in the free vector case. Moreover, there are infinitely many $\hat\CC_{R(j_1,j_2)}$ multiplets in the free vector case with $j_1\ne j_2$, and many have lower spin than the $\hat\CC_{R(j,j)}$ multiplets.}

How should we interpret this result physically? Recall that the free $\CN=2$ Maxwell theory is the IR theory on the Coulomb branch of the MAD theory (e.g., see the Seiberg-Witten based discussion in \cite{Argyres:2015ffa}). In particular, turning on 
\begin{equation}\label{vevMAD}
u:=\langle\CO_{6/5}\rangle\ne0~, 
\end{equation}
where $\CO_{6/5}$ is the dimension $6/5$ chiral ring generator discussed above, we obtain the free $\CN=2$ vector multiplet at long distances. Therefore, we see that the spin thresholds in \eqref{MADgaps} are invariants of these RG flows.
 More generally, we may also consider turning on an $\CN=2$-preserving relevant deformation  
\begin{equation}\label{reldefsMAD}
\delta S=\int d^4x d^4\theta\ h\CO_{6/5}+{\rm h.c.}~,
\end{equation}
where the superspace integral is over all of the $\CN=2$ chiral superspace, and $h$ is a dimension $4/5$ coupling.

For generic $u$ \eqref{vevMAD} and $h$ \eqref{reldefsMAD} the IR theory is again a free vector multiplet \cite{Argyres:2015ffa}. Therefore, we see that for generic $\CN=2$-preserving deformations,\footnote{We are assuming that there are no other $\CN=2$-preserving deformations that consist of turning on vevs for operators other than those parameterizing the Coulomb branch.} the spin thresholds in \eqref{MADgaps} are invariant! 

Only at special co-dimension one points in $u$, $h$ space will the IR theory be different: there we will have massless $N_f=1$ $\CN=2$ SQED. The deep IR then consists of a decoupled vector and a decoupled hypermultiplet of charge $\pm1$ under the gauge group. In this case, the hypermultiplets will give us $\hat\CC_{R(0,0)}$ multiplets for infinitely many $R>0$ (see \eqref{BCspin0} with $n=\tilde n$ to ensure gauge invariance). Then, the IR spin thresholds vanish, and the equality we found above becomes an inequality with smaller spin thresholds in the IR. Of course, we can always decouple the hypermultiplet by turning on an $\CN=2$-preserving mass term, and we are back to the previous situation. 

Given this discussion, one tempting conjecture is that we have the following RG map (from the UV to the IR) for generic $u,h$
\begin{equation}\label{RGmapMAD}
\hat\CC^{\rm MAD}_{R(j_{\rm min}(R),j_{\rm min}(R))} \xrightarrow{\rm RG} \hat\CC^{\rm free\ vec.}_{R(j_{\rm min}(R),j_{\rm min}(R))}~.
\end{equation}
In the case $R=0$, the above is trivially true since we can follow the stress tensor multiplet along the RG flow. For $R>0$, the situation is more subtle. Here we merely note that its proof is not entirely straightforward from the $\CN=1$ UV Lagrangian perspective of \cite{Maruyoshi:2016tqk}.\footnote{The basic reason was explained to us by S.~Giacomelli: consider the superpotential in (2) of \cite{Maruyoshi:2016tqk} and ignore terms involving the decoupling fields $M_1$, $M_3$, and $M_3'$ (they are irrelevant in the flow to the AD theory). Now, turn on $\langle M_5\rangle\ne0$. The $M_5$ multiplet will become an NG multiplet and decouple in the IR. What remains is the superpotential for $\CN=2$ SQCD and $N_f=1$. Therefore we are, in some sense, back to the $\CN=2$ picture of \cite{Argyres:1995xn}. Still, one may hope to use the ideas in \cite{Benvenuti:2017lle,Benvenuti:2017kud} to get a better handle on \eqref{RGmapMAD} by studying the UV chiral multiplets that flow to the IR gauginos.} Finally, let us also note that another question is if every UV $\hat\CC_{R(j,j)}$ MAD multiplet gets mapped to a corresponding IR $\hat\CC_R{(j,j)}$ multiplet in the free vector theory by the RG map (i.e., can we fully embed the UV Schur sector in the IR SCFT?).

In the next subsection we will discuss to what extent the above picture generalizes to various close cousins of the MAD theory.

\subsec{The higher $(A_1, A_{2k})$ SCFTs}
The MAD theory we considered above is the simplest example of an $(A_1, A_{2k})$ SCFT. Indeed, recall from \eqref{equivth}, that the MAD theory is the $(A_1, A_2)$ SCFT. The $k>1$ theories can be roughly thought of as $U(1)^k$ gauge theories with $k$ electric hypermultiplets and $k$  magnetically charged hypermultiplets. It then seems reasonable to expect that aspects of the $k=1$ case generalize. Indeed we will see this is so.

Again following \cite{Foda:2019guo}, the Macdonald index is 
\bea\label{A1A2k}
\CI_M^{(A_1, A_{2k})}(q,T)&=&\sum_{N_1\ge \cdots \ge N_k\ge 0} \frac{q^{N_1^2+\cdots +N_k^2+N_1+\cdots N_k}}{(q)_{N_1-N_2} \cdots (q)_{N_{k-1}-N_k}(q)_{N_k}}T^{N_1+\cdots N_k}\cr&=&\sum_{R=-1}^{\infty} T^{R+1  }  \sum_{N_1\ge \cdots \ge N_k\ge 0\atop N_1+\cdots+N_k=R+1 } \frac{ q^{N_1^2+\cdots +N_k^2+R+1}}{(q)_{N_1-N_2} \cdots (q)_{N_{k-1}-N_k}(q)_{N_k}} ~.
\eea
Let us now define
\be
 \frac{(1-q) }{(q)_{N_1-N_2} \cdots (q)_{N_{k-1}-N_k}(q)_{N_k}} 
 =\sum_{m=0}^\infty f_{N_1, \cdots, N_k; m} q^m~.
\ee
Then, we have
\bea
\CI_M^{(A_1, A_{2k})}(q, T)&=&\sum_{R=-1}^{\infty} \frac{T^{R+1  }}{1-q}  \sum_{N_1\ge \cdots \ge N_k\ge 0\atop N_1+\cdots+N_k=R+1 } \sum_{m=0}^\infty f_{N_1, \cdots, N_k; m} q^{N_1^2+\cdots +N_k^2+R+1 +m} \cr&=&\sum_{R=-1}^{\infty} \sum_j T^{R+1  } \frac{q^{R+2+2j}}{1-q}\sum_{N_1\ge \cdots \ge N_k\ge 0\atop{ N_1+\cdots+N_k=R+1 \atop N_1^2+\cdots +N_k^2+m =2j+1}}
 \sum_{m=0}^\infty f_{N_1, \cdots, N_k; m} ~. 
\eea
Therefore 
 \be\label{inneqA1A2k}
 2j+1=N_1^2+\cdots +N_k^2+   m  \ge 2j_{\rm min}^{(A_1,A_{2k})}(R)+1=\min\left\{N_1^2+\cdots +N_k^2\right\}~,\ \ \ N_1+\cdots N_k=R+1~.
 \ee
As a result, the $\hat {\CC}_{R(j,j)}$ multiplets have spin bounded from below, $j\ge j_{\rm min}^{(A_1,A_{2k})}(R)$, with
\be\label{inneqA1A2k}
j_{\rm min}^{(A_1,A_{2k})}(R)=\min\left\{\frac{N_1^2+\cdots +N_k^2 -1}{2}\Big| N_1+\cdots N_k=R+1, \; N_i \in \mathbb Z_{\ge 0}\right\}   ~.
\ee

Let us now map the inequality in \eqref{inneqA1A2k} onto a generic point of the $k$-dimensional Coulomb branch. To that end, we have, for $i=1,2,\cdots,k$, Schur generators $\lambda_{i,+}^1$ and $\bar\lambda_{i,\dot+}^1$. Let us reinterpret the $N_i$ as the number of $\lambda_{i,+}^1$ factors in the effective IR Schur operator of type $\hat\CC_{R(j_{\rm min}(R),j_{\rm min}(R))}$. So as to minimize spin, such a contribution takes the form $\lambda_{i,+}^1\partial_{+\dot+}\lambda_{i,+}^1\cdots\partial_{+\dot+}^{N_i-1}\lambda_{i,+}^1$. Without loss of generality, we can assume that there is an identical factor involving $\bar\lambda_{i,\dot+}^1$. Therefore, we learn that the contribution from the $i$-th gaugino pair to the spin of the IR Schur operator, $\delta j_{{\rm Schur},i}$, is
\begin{equation}
\delta j_{{\rm Schur},i}={1\over2}(N_i+2(1+2+\cdots+N_i-1))={1\over2}(N_i+N_i(N_i-1))={1\over2}N_i^2~.
\end{equation}
Summing these contributions, the total left spin for the superconformal primary operator accompanying this Schur operator is $(N_1^2+N_2^2+\cdots +N_k^2-1)/2$. Moreover, we must minimize the sum of these contributions subject to the constraint that $N_1+N_2+\cdots+N_k=R+1$ so that the Schur operator has $SU(2)_R$ weight $R+1$.

As a result, we learn that 
\begin{equation}\label{UVIRgenk}
j_{\rm min}^{{\rm (fr. vect.)}^{\times k}}(R)=j_{\rm min}^{(A_1, A_{2k})}(R)~.
\end{equation}
In other words, \eqref{gapEqMAD} generalizes to all $(A_1, A_{2k})$ theories, and $k$ free vectors share all spin thresholds with the $(A_1, A_{2k})$ SCFT!\footnote{Looking closer at \eqref{inneqA1A2k}, one can go slightly further and check that for $R\in \mathbb Z$,
\begin{equation}\label{UVIRgenkForm}
j_{\rm min}^{\rm (fr. vect.)^{\times k}}(R)=j_{\rm min}^{(A_1, A_{2k})}(R)={1\over2}\left(R+(2R+2-k)s-ks^2\right)~, \ \ \ s:={\rm Floor}\left({R+1\over k}\right)~.
\end{equation}
Note that, for $k=1$, this expression reduces to \eqref{MADgaps} and \eqref{SpinFrVect}. Moreover, for $k\to\infty$ and fixed $R$, we get $j_{\rm min}^{\rm (fr. vect.)^{\times k}}(R)=j_{\rm min}^{(A_1, A_{2k})}(R)={R\over2}$ as expected (in the effective Coulomb branch picture, we simply construct the corresponding Schur operator from gauginos without derivatives). For large enough $R$, $j_{\rm min}^{(A_1, A_{2k})}(R)$ can be approximated as $(R+1)^2/(2k)-1/2$, which is obviously a decreasing function  of $k$ for fixed $R$. Moreover, note that the spin thresholds in the $(A_1, A_{2k})$ series are maximized in the MAD theory. This result is in accordance with the minimality of the Schur sector of the MAD theory in section~\ref{minSchur}.  } Moreover, from the point of view of the leading spin operators in the $(A_1, A_{2k})$ theory, we see that a natural interpretation of the $k$ different particle species in \cite{Foda:2019guo} is as the $k$ different effective gaugino pairs.\footnote{Although note that the leading spin operators for all $(A_1, A_{2k})$ theories are unique for any $R$. On the other hand, in the IR theory the multiplicity can be higher (this follows from the decoupled nature of the IR). } In particular, we see that the spin thresholds are preserved under generic $\CN=2$-preserving deformations of the $(A_1, A_{2k})$ theory, 
\begin{equation}\label{reldefsA1A2k}
u_i:=\langle\CO_{\Delta_i}\rangle\ne0~, \ \ \ \delta S=\int d^4x d^4\theta\ \left(\sum_{i=1}^kh_i\CO_{\Delta_i}\right)+{\rm h.c.}~,
\end{equation}
where $\Delta_i:=2-2(1+i)/(3+2k)$ are the scaling dimensions of the corresponding $k$ generators, $\CO_{\Delta_i}$ ($i=1,2,\cdots,k$), of the $(A_1, A_{2k})$ chiral ring, and the $h_i$ are dimension $2(1+i)/(3+2k)$ relevant couplings.

In light of this discussion, it would be interesting to understand if, under generic RG flows, we have the following map of operators
\begin{equation}\label{RGmap2k}
\hat\CC_{R(j_{\rm min}(R),j_{\rm min}(R))}^{(A_1, A_{2k})}\xrightarrow{\rm RG}\hat\CC_{R(j_{\rm min}(R),j_{\rm min}(R))}^{{\rm (fr. vect.)}^{\times k}}~,
\end{equation}
Note that the target of this map has an ambiguity since there are generally multiple such operators in the IR. Therefore, this map should be made more precise for $k>1$.

Somewhat more generally, we have Coulomb branch RG flows of the form (e.g., see \cite{Xie:2013jc})
\begin{equation}
(A_1, A_{2k})\to(A_1, A_{2(k-n)})+{{\rm (fr. vect.)}^{\times n}}~,
\end{equation}
at special loci of the $(A_1, A_{2k})$ moduli space (i.e., by appropriately tuning the deformations in \eqref{reldefsA1A2k}). These flows can be read off from the corresponding Seiberg-Witten curves as in \cite{Xie:2013jc}. Our discussion above also shows that
\begin{equation}
j_{\rm min}^{(A_1, A_{2(k-n)})+{\rm (fr. vect.)^{\times n}}}(R)=j_{\rm min}^{(A_1, A_{2k})}(R)~.
\end{equation}
It is then natural to wonder if there is a map of operators\footnote{Perhaps one can argue that this should hold since our operators ultimately appear in the $n$-fold product of stress tensor multiplets and are therefore, in some sense, universal.}
\begin{equation}
\hat\CC_{R(j_{\rm min}(R),j_{\rm min}(R))}^{(A_1, A_{2k})}\xrightarrow{
\rm RG}\hat\CC_{R(j_{\rm min}(R),j_{\rm min}(R))}^{(A_1, A_{2(k-n)})+{{\rm (fr. vect.)}^{\times n}}}~.
\end{equation}

In the next section we study theories with Higgs branches. Unlike the theories considered so far, these theories can have vanishing spin thresholds.

\newsec{Spin thresholds and theories with Higgs branches}\label{genspin}
In this section, we begin to tackle the question of, to what extent, \eqref{gapEqMAD} and \eqref{UVIRgenk} generalize. We do not have a complete picture yet, but we will show that such simple relations generally need to be modified when a theory has a Higgs branch (as we will see, these theories necessarily have infinitely many $\hat\CC_{R(j,j)}$ multiplets with low spin). However, it is possible that with some caveats (e.g., excluding cases with irrelevant gauging in the IR) interesting general UV/IR spin threshold inequalities can be established for flows onto a pure Coulomb branch (here \lq\lq pure" Coulomb branch means that the Coulomb branch only consists of free vectors at generic points).\footnote{Even for flows onto the Higgs branch, if we are willing to take the SCFT on the Higgs branch and give masses to the free hypermultiplets, then it seems plausible that we can, through a sequence of flows, end up with a theory that has all spin thresholds larger than in the UV (essentially because we decouple more and more \lq\lq matter" and end up with theories that either are trivial or else have no Higgs branch).}

\subsec{The free hypermultiplet and SCFTs with a Higgs branch}
In section \ref{A1A2spin}, we saw that the MAD theory has spin thresholds in its $\hat\CC_{R(j,j)}$ spectrum that grow quadratically with $R$. On the other hand, for theories with Higgs branches, we will see that there are infinitely many low-spin $\hat\CC_{R(j,j)}$ multiplets. Therefore, in such theories, the RG map \eqref{RGmapMAD} is necessarily more complicated (e.g., as in the case of the $(A_1, A_{2n+1})$ SCFTs).

To that end, consider the simplest SCFT with a Higgs branch: the free hypermultiplet. The degrees of freedom can be packaged into $\CN=1$ (anti-)chiral superfields as follows
\begin{equation}\label{qtq}
Q^i:=\begin{pmatrix}
Q\\ \tilde Q^{\dagger}
\end{pmatrix}~,
\ \ \ Q^{\dagger i}:=\begin{pmatrix} \tilde Q & -Q^{\dagger}
\end{pmatrix}~,
\end{equation}
where $i=1,2$ is a fundamental index of $SU(2)_R$ (i.e., the free hypermultiplet is a $\hat\CB_{1\over2}$ multiplet). The statement that the theory has a Higgs branch is equivalent to the existence of an infinite set of $\hat\CB_R$ multiplets. These multiplets have primaries
\begin{equation}
\hat\CB_{R}\ni Q^{(i_1}Q^{i_2}\cdots Q^{i_n}Q^{\dagger j_1}Q^{\dagger j_2}\cdots Q^{\dagger j_{\tilde n})}~, \ \ \ n+\tilde n=2R~, 
\end{equation}
where $n$ and $\tilde n$ are the number of $Q$ and $\tilde Q$ generators respectively, and \lq\lq$(\cdots)$" denotes symmetrization of the enclosed indices. Since there are no relations amongst these generators, we see that the Higgs branch is $\CM_{H}=\mathbb{C}^2$.

What about more general operators associated with the existence of a Higgs branch and with locality? Locality implies the existence of a stress tensor multiplet with primary
\begin{equation}
\hat\CC_{0(0,0)}\ni \epsilon_{ij}Q^iQ^{\dagger j}~.
\end{equation}
Given this primary, we can also form the normal-ordered primary
\begin{equation}\label{BCspin0}
\hat\CC_{R(0,0)}\ni \epsilon_{ij}Q^{(i_1}Q^{i_2}\cdots Q^{i_n}Q^{\dagger j_1}Q^{\dagger j_2}\cdots Q^{\dagger j_{\tilde n})}Q^iQ^{\dagger j}~,\ \ \ n+\tilde n=2R~.
\end{equation}
Therefore, we have a scalar $\hat\CC_{R(0,0)}$ multiplet for any possible $SU(2)_R$ weight. Similarly, we can construct $\hat\CC_{R-1(1/2,1/2)}$ multiplets. For example, we have at least one primary of the form
\begin{equation}\label{BCspinhalf}
\hat\CC_{R-1(1/2,1/2)}\ni \epsilon_{ij}Q^{(i_1}Q^{i_2}\cdots Q^{i_n}Q^{\dagger j_1}Q^{\dagger j_2}\cdots Q^{\dagger j_{\tilde n})}Q^i\partial_{\alpha\dot\alpha}Q^{\dagger j}+\cdots~,\ \ \ n+\tilde n=2(R-1)~.
\end{equation}
In other words, the free hypermultiplet's $\hat\CC_{R(j,j)}$ spectrum has, for any $SU(2)_R$ weight, multiplets of spin zero and total spin one (in particular, all spin thresholds vanish). Note that under the 4D/2D map both \eqref{BCspin0} and \eqref{BCspinhalf} contribute chiral algebra states with $h=\Delta-R=2+R$.

In the case of the free hypermultiplet, the theory has no Coulomb branch and so a generalization of \eqref{RGmapMAD} and \eqref{UVIRgenk} doesn't immediately make sense. However, we will see that more general theories with Higgs branches and Coulomb branches also have infinitely many $\hat\CC_{R(j,j)}$ multiplets with total spin less than or equal to one. Therefore, \eqref{RGmapMAD} and \eqref{UVIRgenk} do not directly apply in these cases.

Before we consider such more general theories, let us note that the free hypermultiplet has an $SU(2)$ flavor symmetry (here $Q$ and $\tilde Q$ form a flavor doublet). Therefore, we have a $\hat\CB_1$ multiplet containing the corresponding Noether current and transforming in the adjoint. The primary of this multiplet takes the form
\begin{equation}\label{B1}
\hat\CB_1\ni \begin{pmatrix}
Q^{(i_1}Q^{j_1)}\\ Q^{(i_1}Q^{\dagger j_1)}\\ Q^{\dagger(i_1}Q^{\dagger j_1)}
\end{pmatrix}~.
\end{equation}
Our above discussion shows that there is also an $SU(2)$-adjoint valued $\hat\CC_{1(0,0)}$ multiplet with primary
\begin{equation}\label{C100fh}
\hat\CC_{1(0,0)}\ni \begin{pmatrix}
\epsilon_{ij}Q^{(i_1}Q^{j_1)}Q^i\tilde Q^{\dagger j}\\ \epsilon_{ij}Q^{(i_1}Q^{\dagger j_1)}Q^i\tilde Q^{\dagger j}\\ \epsilon_{ij}Q^{\dagger(i_1}Q^{\dagger j_1)}Q^i\tilde Q^{\dagger j}
\end{pmatrix}~.
\end{equation}
We will see that any local 4D $\CN=2$ SCFT with a locally realized flavor symmetry has an adjoint-valued $\hat\CC_{1(0,0)}$ multiplet as well. Moreover, as discussed more above, we also have a $\hat\CC_{0(1/2,1/2)}$ multiplet transforming in the adjoint (this statement is consistent with the fact that the hypermultiplet is free).

Now let us consider an abstract local interacting 4D $\CN=2$ SCFT with some locally realized flavor symmetry.  OPE selection rules dictate that (e.g., see \cite{Kiyoshige:2018wol})
\begin{equation}\label{C0B1}
\hat\CC_{0(0,0)}\times\hat\CB_1\sim\hat\CB_1+\sum_{\ell=0}^{\infty}\left[\hat\CC_{1(\ell/2,\ell/2)}+\hat\CC_{0((\ell+1)/2,(\ell+1)/2)}\right]+\cdots~,
\end{equation}
where the ellipses include non-Schur multiplets. Since the product of Schur operators on the left hand side has $h=\Delta-R=3$, we see that both $\hat\CC_{1(0,0)}$ and $\hat\CC_{0(1/2,1/2)}$ can contribute to the 2D normal-ordered product of the holomorphic stress tensor and affine current, $TJ^a$ (here $a$ is an adjoint index). Since the theory is interacting, $\hat\CC_{0(1/2,1/2)}$ is absent. In general, there can be other OPEs that give rise to adjoint-valued $\hat\CC_{1(0,0)}$ multiplets. For example, the $\hat\CB_1\times\hat\CB_1$ OPE can also be a source of such multiplets. The relevant selection rule in this case is\cite{Nirschl:2004pa}
\begin{equation}\label{B1B1}
\hat\CB_1\times\hat\CB_1\sim1+\hat\CB_1+\hat\CB_2+\sum_{\ell=0}^{\infty}\left(\hat\CC_{0(\ell/2,\ell/2)}+\hat\CC_{1(\ell/2,\ell/2)}\right)+\cdots~.
\end{equation}
We leave the derivation of a $\hat\CC_{1(0,0)}$ multiplet in this channel to Appendix \ref{appA}. In some theories (e.g., the $(A_1, A_3)$ and $(A_1, D_4)$ SCFTs), \eqref{C0B1} and \eqref{B1B1} do not give rise to more than one independent $\hat\CC_{1(0,0)}$ multiplet in total.\footnote{For the $(A_1, A_3)$ and $(A_1, D_4)$ SCFTs, one can check that there is a null relation involving the non-null $J^aT$, $\partial^2J^a$, and $f^{abc}J_b\partial J_c$ operators (note that here $T$ is the Sugawara stress tensor).} In these cases, the above mechanisms produce only a single $\hat\CC_{1(0,0)}$ multiplet.

How do we see that at least one adjoint-valued $\hat\CC_{1(0,0)}$ multiplet must be present? Recall that there is a constant $\kappa$ such that $TJ^a+\kappa \partial^2J^a$ is an $sl(2,\mathbb{R})$ primary (this statement follows from acting with $L_1$ on the corresponding state and using the Virasoro and affine Kac-Moody commutation relations; it holds regardless of other operator relations). This primary must correspond to an adjoint-valued $\hat\CC_{1(0,0)}$ multiplet, and this multiplet is absent if and only if $TJ^a+\kappa \partial^2J^a$ is null in the chiral algebra. Thinking of $J^a$ as an $h=1$ Virasoro primary, we see that $TJ^a +\kappa \partial^2J^a$ can be null only if (e.g., see (7.26) of \cite{DiFrancesco:1997nk})
\begin{equation}\label{descnull}
1={1\over16}(5-c_{2d}-\sqrt{(1-c_{2d})(25-c_{2d})})\ \ \ {\rm or}\ \ \ 1={1\over16}(5-c_{2d}+\sqrt{(1-c_{2d})(25-c_{2d})})~,
\end{equation}
and so $c_{2d}=-2$. However, this corresponds to the 4D central charge of a free vector multiplet, $c=1/6$ (this particular theory anyway has no flavor symmetry; the $h=1$ states correspond to 4D gauginos). Since it has less than the central charge of the MAD theory, the results of \cite{Liendo:2015ofa} guarantee the theory is free. As a result, we learn that

\medskip
\noindent
{\bf Claim 2:} Any local interacting 4D $\CN=2$ SCFT with a locally realized flavor symmetry has an adjoint-valued $\hat\CC_{1(0,0)}$ multiplet. In particular, such theories cannot have non-vanishing spin thresholds for all $R>0$. 

\medskip
\noindent

Let us now consider more general Higgs branch operators. Since there is a Higgs branch, there is an infinite set of corresponding $\hat\CB_R$ operators with arbitrarily large $R$ whose Schur operators can acquire a vev. In this case \eqref{C0B1} becomes \cite{Kiyoshige:2018wol}
\begin{equation}
\hat\CC_{0(0,0)}\times\hat\CB_R\sim \hat\CB_R+\sum_{\ell=0}^{\infty}\left(\hat\CC_{R(\ell/2,\ell/2)}+\hat\CC_{R-1(\ell/2,\ell/2)}\right)+\cdots~, \ \ \ R>1~,
\end{equation}
where non-Schur contributions are included in the ellipses. Since the product of Schur operators on the left hand side has $h=\Delta-R=2+R$, we see that both $\hat\CC_{R(0,0)}$ and $\hat\CC_{R-1(1/2,1/2)}$ (each having Schur operators with $h=2+R$) can contribute to the 2D normal-ordered product $TB$ (where $B$ is the 2D chiral algebra generator corresponding to the $\hat\CB_R$ multiplet). Again, we can have other sources of $\hat\CC_{R(0,0)}$ and $\hat\CC_{R-1(1/2,1/2)}$ multiplets arising from OPEs like $\hat\CB_R\times\hat\CB_R$ \cite{Nirschl:2004pa}.

Similar reasoning to that employed in \eqref{descnull} shows that there is a $\kappa'$ such that $TB+\kappa'\partial^2T$ is an $sl(2,\mathbb{R})$ primary. It vanishes only if 
\begin{equation}
R={1\over16}(5-c_{2d}-\sqrt{(1-c_{2d})(25-c_{2d})})\ \ \ {\rm or}\ \ \ R={1\over16}(5-c_{2d}+\sqrt{(1-c_{2d})(25-c_{2d})})~.
\end{equation}
Clearly, for fixed $c$ and $c_{2d}$, there are infinitely many $R$ such that the normal-ordered product is non-vanishing. Unlike the $R=1$ case, here we cannot immediately conclude that $\hat\CC_{R-1(1/2,1/2)}$ is vanishing. We therefore learn that:

\medskip
\noindent
{\bf Claim 3:} Any local interacting 4D $\CN=2$ SCFT with a Higgs branch has, for infinitely many values of $R\in\mathbb{Z}$, at least one $\hat\CC_{R(0,0)}$ or $\hat\CC_{R(1/2,1/2)}$ multiplet.\footnote{As mentioned above, the theory has at least one Schur operator in a $\hat\CB_R$ multiplet that can acquire a vev (this statement holds even if there are Higgs branch chiral ring relations). Taking even-order products of this operator with itself gives us infinitely many $\hat\CB_{2nR}$ multiplets. Clearly $2nR\in\mathbb{Z}$.}

\medskip
\noindent
Therefore, we learn that any theory with a Higgs branch will necessarily have infinitely many $\hat\CC_{R(j,j)}$ multiplets that decouple (or flow to non-Schur multiplets) on an RG flow to a generic point on the pure Coulomb branch.\footnote{This statement follows from the fact that Poincar\'e symmetry is unbroken. Note that if the Coulomb branch has complex dimension $n$, it has Schur operators $\lambda_{i,+}^1$ and $\bar\lambda_{i,\dot+}^1$ (with $i=1,2,\cdots,n$). This spectrum implies that the minimal spin IR $\hat\CC_{R(j,j)}$ multiplet has $j_{\rm min}(R)\ge{R\over2}$. Therefore, all UV $\hat\CC_{R(0,0)}$ or $\hat\CC_{R(1/2,1/2)}$ multiplets with $R\ge2$ will decouple (or flow to non-Schur multiplets).} 
In particular, the small (or vanishing) spin thresholds for $R\ge2$ will all grow in the IR. 

In this light, the behavior of the MAD theory in \eqref{gapEqMAD} and of the higher $(A_1, A_{2k})$ SCFTs in \eqref{UVIRgenk} is special. More generally, it may be possible to show that the $j^{\rm IR}_{\rm min}(R)\ge j_{
\rm min}^{\rm UV}(R)$ spin thresholds must increase in RG flows to a pure Coulomb branch.\footnote{We do not have any counterexamples to this statement for RG flows to generic points on the Coulomb branch.  As we have already indicated above, in cases where we have RG flows featuring irrelevant gauging of IR flavor symmetries (e.g., as in \cite{Argyres:2012fu} or as in massless SQED points on more general Coulomb branches), we need to be more careful and potentially refine our statement.} Note that on flows to the Higgs branch the opposite is trivially true: \eqref{BCspin0} implies that $j^{IR}
_{\rm min}(R)\le j_{\rm min}^{UV}(R)$. But it may be possible to reverse the inequality by turning on mass terms.

\newsec{Minimality of the MAD Schur sector}\label{minSchur}
In this section we turn our attention to the minimality of the MAD Schur sector. In particular, we will prove that:

\medskip
\noindent
{\bf Claim 4:} Among all local unitary 4D $\CN=2$ SCFTs with only bosonic Schur operators, the MAD theory has the smallest Schur index. In other words, for every $h=\Delta-R$, the corresponding coefficient in the Schur index is smallest.\footnote{The only theory we are aware of with smaller Schur index (i.e., each coefficient is less than or equal to the corresponding coefficient of the $(A_1, A_2)$ Schur index, and infinitely many are smaller) is the $(3,2)$ theory (i.e., the diagonal conformal $su(2)$ gauging of three copies of the $(A_1, A_3)$ SCFT). Note that its chiral algebra has, in addition to the energy momentum tensor, two fermionic generators at $h=4$, and the Schur sector of this theory is intimately related to that of $\CN=4$ super-Yang-Mills theory \cite{Buican:2016arp,Buican:2020moo} (see also related index studies in \cite{Kang:2021lic}).
}

\medskip
\noindent
The list of theories with only bosonic Schur operators is vast (e.g., to our knowledge it includes all isolated $(G,G')$ theories for which the Schur sector is known and all isolated purely $\CN=2$ class $\CS$ theories for which the Schur sector is known). More interestingly, the above claim also implies the following: 

\medskip
\noindent
{\bf Corollary 5:} Among all local unitary 4D $\CN=2$ SCFTs (including those with fermionic Schur operators), the MAD SCFT has the smallest number of Schur operators for every $h=\Delta-R$.

\medskip
\noindent
This result follows trivially from Claim 4 for theories with only bosonic Schur operators, because the Schur index then counts, for each $h$, the number of operators at that level. For theories with fermionic generators, the claim also follows trivially if each coefficient in the corresponding Schur index is not smaller than the corresponding coefficient in the MAD Schur index.

More generally, Corollary 5 follows from Claim 4, because we will establish Claim 4 by showing that:

\medskip
\noindent
{\bf Claim 6:} The MAD SCFT has the smallest associated Virasoro vacuum character of any local unitary 4D SCFT (i.e., for any $h$, the corresponding entry in the vacuum character is smallest).\footnote{Locality implies the existence of an energy momentum tensor and hence, by the 4D/2D map, a corresponding two-dimensional holomorphic stress tensor and associated Virasoro vacuum module.}

\medskip
\noindent
In particular, if another SCFT has additional chiral algebra generators beyond the stress tensor, these will not lower the number of operators at level $h$ beyond those counted by the Virasoro vacuum character. Indeed, the only way to remove Virasoro vacuum operators in our counting is for them to be involved in null relations with non-Virasoro vacuum operators (but then we must count the corresponding non-Virasoro vacuum operators with the same $h$). This statement holds even if $T$ is composite (e.g., as in the Sugawara case), since we can always define a Virasoro vacuum module (the affine currents themselves are not part of this Virasoro vacuum module).

In the next subsection we prove Claim 6 (and therefore Claim 4 and Corollary 5) by using the results of \cite{Welsh:2002jq} to show that the Lee-Yang Virasoro vacuum character (i.e., the vacuum character associated with the MAD theory) is the smallest among the vacuum characters of all non-unitary minimal models.\footnote{Strictly speaking, this is a stronger statement than what we need to prove Claim 6. Indeed, certain non-unitary minimal models have central charges that cannot correspond to any unitary 4D SCFTs (e.g., the $(3,5)$ minimal model gives rise to a $c$ in 4D that violates the bounds of \cite{Liendo:2015ofa}; note that $c$ in this case does not correspond to free 4D fields).} While we believe the unitary theories also have larger vacuum characters than Lee-Yang (and have checked this numerically to high order), this point is irrelevant for us since unitary 2D theories can, at best, correspond to non-unitary 4D $\CN=2$ SCFTs.

\subsec{Minimality of the Lee-Yang vacuum character}
In this subsection, we use the results of \cite{Welsh:2002jq} to argue that the Lee-Yang vacuum character is smallest among all Virasoro vacuum characters for non-unitary 2D theories.  

For generic $c$ and $c_{2d}$, the Virasoro vacuum module is obtained by quotienting the Verma module by the $L_{-1}|0\rangle$ null vector
\begin{equation}\label{genchi0}
\chi_{1,1}^{c_{2d,{\rm generic}}}(q)={1\over\prod_{i=2}(1-q^i)}~.
\end{equation}
Here we have normalized the identity operator's contribution to unity.

On the other hand, as is well known from the Kac formula
\begin{equation}
0=h_{r,s}={((m+1)r-ms)^2-1\over 4m(m+1)}~,\ \ \ c_{2d}=1-{6\over m(m+1)}~,
\end{equation}
we have a non-trivial null vector in the vacuum module (i.e., one with $rs>1$) after quotienting out by $L_{-1}|0\rangle$ only if
\begin{equation}
c_{2d}=1-6{(k-k')^2\over kk'}~,
\end{equation} 
where $k=r\pm1$, $k'=s\pm1$, and we require that, for a particular choice of the sign to be valid, $k,k'>0$. In writing down $c_{2d}$ we can divide out any common factors of $k$ and $k'$ and write the result in terms of $p$ and $p'$ with
\begin{equation}\label{minc}
c_{2d}=1-6{(p-p')^2\over pp'}~,\ \ \ \gcd(p,p')=1~,\ \ \ p'> p~.
\end{equation} 
The cases $p\ge2$ and $p=1$ are qualitatively different. For $p=1$, the vacuum module we get by quotienting out by $L_{-1}|0\rangle$ is irreducible and has the form in \eqref{genchi0} \cite{Gaberdiel:1996kx}. Therefore, for our purposes, we may treat the $p=1$ case in the same way as the generic central charge case. On the other hand, for $p\ge2$, additional null vectors appear.

The Lee-Yang minimal model has $p=2$ and $p'=5$, and so it is of this latter class. Therefore, we have
\begin{equation}\label{LYgeneric}
\chi_{1,1}^{(2,5)}<\chi_{1,1}^{c_{2d, {\rm generic}}}~,
\end{equation}
We use \lq\lq$<$" to indicate that, at each order in $q$, the coefficient of the Lee-Yang vacuum character is less than or equal to the corresponding quantity for the generic central charge theory, and that there is at least one power of $q$ for which the Lee-Yang coefficient is smaller (e.g., $q^4$, where the first Lee-Yang null state enters).

In what follows, we would like to extend \eqref{LYgeneric} by replacing the right hand side with the vacuum characters of any non-unitary minimal model. As discussed above, any unitary minimal models will necessarily correspond to 4D central charge $c\le0$ and so will be non-unitary in the higher dimension.

To accomplish this task, our strategy is to use the results of \cite{Welsh:2002jq}. In particular, we start by rewriting $p'/p$ as follows
 \be\label{condfrac}
 \frac{p'}{p} =c_0+\frac{1}{c_1+\frac{1}{c_2+\frac{1}{ {\vdots\atop c_{n-1}+ \frac{1}{c_n} }   }}}~, \qquad c_0~, \cdots, c_{n-1}\ge 1~, \; c_n\ge 2~.
 \ee
Then,  $[c_0,c_1, \cdots, c_n]$ is said to represent the continued fraction of $p'/p$, and $n$ is called the height of $p'/p$.  We are interested in the vacuum character of the minimal model $\chi^{[c_0,c_1, \cdots, c_n]}:=\chi^{(p,p')} _{1,1} $. 

As we will soon see, the formulas in \cite{Welsh:2002jq} are useful, because they have certain manifest positivity properties that we will make use of. In particular, consider (1.16) of \cite{Welsh:2002jq}
 \be\label{recurch}
 \chi^{(p,p')}_{r,s}:=\chi_{r,s}^{[c_0,c_1,\cdots,c_n]}=F^{(p,p')}_{r,s}+
 \begin{cases}
 \chi^{(\hat p,\hat p')}_{\hat r,\hat s}~, & {\rm if}\ \eta(s)=\tilde\eta(r)\ {\rm and}\ \hat s,\hat r\ne0~, \\
 0 ~, & \text{otherwise~,}
 \end{cases}
 \ee
 where $F$ is given by an infinite sum, and each summand is given  by some  power of $q$  multiplied by $q$-Pochhammer symbols and $q$-binomial coefficients. An explicit expression for $F$ is rather complicated in general, but it can be determined following the prescription in  \cite{Welsh:2002jq} (we refer the reader to this text for a definition of $\eta$ and $\tilde\eta$ above). Here, we only need to know that $F$ is positive: namely, as a series in $q$, it only has positive coefficients.  Moreover, we are only interested in the vacuum module, $r=s=1$, and so the discussion in \cite{Welsh:2002jq} implies that $\hat r=\hat s=1$. Furthermore, it turns out that $\hat p'/\hat p$ has the same continued fraction as $p'/p$ except for the removal of the last entry (see (1.34) and (1.35) of \cite{Welsh:2002jq}).

It also turns out that the first condition in \eqref{recurch} is satisfied as long as $\hat p'/\hat p$ is not an integer. As a result, we have the following recursion relation for  the vacuum character 
\be
\chi^{[c_0, c_1, \cdots, c_n]}_{1,1}
=\text{positive series}+\chi_{1,1}^{[c_0, c_1, \cdots, c_{n-1}]}~, \qquad n\ge 2~,
\ee
if $c_{n-1}>1$. In the case that $c_{n-1}=1$, we should  replace  $[c_0, c_1, \cdots, c_{n-1}=1]$ 
with $[c_0, c_1, \cdots, c_{n-2}+1]$; these expressions correspond to the same fraction $\hat p'/\hat p$, but only the latter is an allowed continued fraction in the conventions of \eqref{condfrac}. Accordingly, we have
\be\label{recur2}
\chi^{[c_0, c_1, \cdots, c_{n-1}=1, c_n]}_{1,1}
=\text{positive series}+\chi^{[c_0, c_1, \cdots, c_{n-2}+1]}_{1,1}~, \qquad n\ge 3~.
\ee

These two recursion relations enable  us to compute the character recursively. At each step, the  vacuum character of a minimal model reduces to the sum of a positive series and the vacuum character of another minimal model with smaller height. By induction, we thus have
\be\label{gtchi}
  \chi^{[c_0, c_1', \cdots, c_n']}_{1,1}
  > \chi^{[c_0, c_1 ]}_{1,1}~, \qquad
  \text{ or }\qquad
    \chi^{[c_0, 1,c_2', \cdots, c_n']}_{1,1} 
    > \chi^{[c_0, 1,c_2 ]}_{1,1}~, \qquad c_1~,\ c_2\ge 2~.
\ee 
Note that $c_{1,2}$ may be different from $c'_{1,2}$ due to equation \eqref{recur2}. We leave the second choice in \eqref{gtchi} as $[c_0, 1,c_2 ]$, because otherwise \eqref{recur2} gives the label $[c_0+1]$ (which results in $\hat p'/\hat p\in\mathbb{Z}$). Therefore, we only need to consider the case $[c_0,c_1]$ and $[c_0,1,c_2]$. In these two terminal cases, it turns out that the $F$ in \eqref{recurch} is considerably simpler. 
 
 \medskip
 \noindent
{\bf Case 1: $\bm{ [c_0,c_1]}$.} Here we have $p'/p=c_0+1/c_1$ with $c_0\ge1$ and $c_1\ge2$. We introduce $h:=c_0-1$ and $k:=c_1-2$ such that $p'/p=h+1+1/(k+2)$. Note that $h,k$ are non-negative integers.   
  
In this case, $\chi_{1,1}^{[c_1,c_2]}=\chi_{1,1}^{[h+1,k+2]}=F$, and we can rewrite $F$---whose general expression is given in (1.17) of \cite{Welsh:2002jq}---as follows\footnote{The components of our $\bm m$ differ from those in \cite{Welsh:2002jq} as follows $m_{i}={1\over2}m^{\rm there}_{t_1+i}$, our $\bm C$ is $\overline{\bm{C}}$ of \cite{Welsh:2002jq}, and we will explain the difference between our $\bm{n}_{L,R}$ and $\tilde{\bm{n}}$ in \cite{Welsh:2002jq} below. \label{dictionaryF}}
\be\label{chivacn}
  \chi^{[h+1, k+2]} =\sum_{  \bm m \in \mathbb Z_{\ge 0}^k \atop \bm n \in \mathbb Z_{\ge 0}^h}
 q^{\bm m \bm C \bm m^T+\bm n_L\bm  B \bm n_R^T}
 \times
   \frac{1}{(q)_{2m_{ 1}}}\prod_{i=1}^h \frac{1}{(q)_{n_i}}
  \prod_{i= 2}^ {k  } \Big[{  m_{i-1}+m_{i+1} \atop 2 m_i}\Big]_q~.
\ee
 In the above equation, we  have used the $q$-binomial coefficient, which is defined as
 \be\label{qBinomial}
 \Big[{n \atop m}\Big]_q:=
 \frac{(q)_n}{(q)_m(q)_{n-m}}~,  \qquad\text{ if } \quad 0 \le m \le n~,
\ee
and vanishes otherwise. The $q$-Pochhammer symbol is defined in \eqref{qpoch}. In the limit $q\to 1$, \eqref{qBinomial} reduces to the standard binomial  coefficients (i.e., $\big({n \atop m}\big)=n!/m!/(n-m)!$).
 Note that the  $q$-binomial coefficients are series in $q$ with positive coefficients. Positivity can be seen from the following identities\footnote{See, e.g., \url{https://en.wikipedia.org/wiki/Q-Pochhammer_symbol}}
 \be 
   \Big[{n \atop m}\Big]_q=  \Big[{n-1 \atop m-1}\Big]_q+ q^m \Big[{n-1 \atop m}\Big]_q ~.
 \ee
 
In writing \eqref{chivacn}, we sum over the $k$-dimensional vector with non-negative integer entries, $\bm  m=(m_1, \cdots, m_k) $, and we set $m_{k+1}=0$; $\bm C$ is a $k\times k$ matrix defined by 
 \be
( \bm C)_{ab}=2 \delta_{a,b} -\delta_{a-b,1}-\delta_{b-a,1}~, \qquad a,b=1, \cdots, k~,
 \ee
 so that the non-vanishing elements only sit  along the diagonal and next-to-diagonal directions. We also sum over the entries in the vector $\bm  n=(n_1, \cdots, n_h) $, and $\bm B$ is an $h\times h$ matrix defined as 
 \be
( \bm B)_{ab}=\min(a,b)~, \qquad a,b=1~, \cdots~, h~.
 \ee
 The two $h$-dimensional vectors $\bm n_L, \bm n_R$ are given by
 \be
  \bm  n_L =(n_1, \cdots, n_{h-1}, n_h+m_1)  ~, \qquad 
  \bm   n_R =(n_1, \cdots, n_{h-1}, n_h+m_1+1) ~.
 \ee
 In the case that $k=0$, we set $m_1=0$. To make closer contact with \cite{Welsh:2002jq}, the $\bm B$-dependent exponent of $q$ in \eqref{chivacn} can also be written as
 \be\label{ndictionary}
\bm n_L\bm  B \bm n_R^T=\tilde {\bm n} \bm  B \tilde {\bm n}^T-h/4, 
 \qquad \tilde {\bm n}=\frac{\bm n_L+\bm n_R}{2}=\left(n_1, \cdots, n_{h-1}, n_h+m_1+\frac12\right) ~.
 \ee
 Indeed, we can recover (1.17) of \cite{Welsh:2002jq} using the dictionary in footnote \ref{dictionaryF}, \eqref{ndictionary}, and the fact that $\gamma=-h$ (this latter relation follows from applying the discussion around (1.25)-(1.27) of \cite{Welsh:2002jq}).\footnote{As a check on this discussion, let us rewrite \eqref{ndictionary} as follows:
 \be\label{chiNn1}
\bm n_L\bm  B \bm n_R^T=\tilde {\bm n} \bm  B \tilde {\bm n}^T-h/4=\sum_{j=1}^h (\tilde N_j^2+\tilde N_j)~, \qquad \tilde N_j
=  n_{Lj}+\cdots +  n_{Lh}~.
 \ee
 or
    \be\label{chiNn}
\bm n_L\bm  B \bm n_R^T=\tilde {\bm n} \bm  B \tilde {\bm n}^T-h/4=\sum_{j=1}^h (  (N_j+m_1)^2+  N_j+m_1)~, \qquad 
 N_j
=  n_{ j}+\cdots +  n_{ h}~.
 \ee
Plugging the above into \ref{chivacn} and setting $k=0$, the formula above becomes
   \be\label{A1A2ks}
\chi^{(2,2h+3)}(q )=\sum_{N_1\ge \cdots \ge N_h\ge 0} \frac{q^{N_1^2+\cdots +N_h^2+N_1+\cdots N_h}}{(q)_{N_1-N_2} \cdots (q)_{N_{h-1}-N_h}(q)_{N_h}}~,
\ee
which is a known expression for the $(2,2h+3)$ Virasoro vacuum modules (e.g., see the recent discussion in \cite{Foda:2019guo} or \eqref{A1A2k} with $T\to 1$). Setting $h=0$ and allowing any $k\ge1$, we recover the \lq\lq fermionic" unitary minimal model expression of \cite{Melzer:1993zk}
 \be\label{chivacnU}
  \chi ^{(k+2,k+3)}=\sum_{  \bm m \in \mathbb Z_{\ge 0}^k}
  \frac{1}{(q)_{2m_1}}q^{\bm m \bm C\bm  m^T}
  \prod_{i=2}^k \Big[{  m_{i-1}+m_{i+1} \atop 2 m_i}\Big]_q~.
 \ee
 }

\medskip
\noindent
{\bf Case 2: $\bm{ [c_0,1,c_2]}$.} In the second case, $p'/p=c_0+1/(1+1/c_2)$. We introduce $h=c_0-1, k=c_2-1$ such that $p'/p=h+1+1/(1+1/(k+1))$. Note that $h,k$ are again non-negative integers.    
   In this case, the vacuum character turns out to be given by 
   \be\label{chivacnii}
  \chi^{[h+1, 1, k+1]} =\sum_{  \bm m \in \mathbb Z_{\ge 0}^k \atop \bm n \in \mathbb Z_{\ge 0}^h}
 q^{\bm m \tilde{\bm C} \bm m^T+\bm n_L\bm  B \bm n_R^T+m_1}
 \times
   \frac{1}{(q)_{2m_{ 1}}}\prod_{i=1}^h \frac{1}{(q)_{n_i}}
  \prod_{i= 2}^ {k  } \Big[{  m_{i-1}+m_{i+1} \atop 2 m_i}\Big]_q~.
 \ee
Except for the matrix $\tilde{\bm C}$ and the explicit appearance of  $m_1$ in the exponent of $q$, this formula is very similar to \eqref{chivacn}. This matrix is defined as
\be
(\tilde{\bm C} )_{ab}=2 \delta_{a,b} -\delta_{a-b,1}-\delta_{b-a,1}~, \qquad a =2, \cdots, k~,\qquad
b=1, \cdots, k~,
 \ee
 and 
 \be
(\tilde  {\bm C} )_{11}=1~, \qquad  (\tilde{\bm C} )_{12}=1~, \qquad (\tilde{\bm C} )_{13}= \cdots  (\tilde{\bm C}) _{1k}=0~.
 \ee
Therefore, $\tilde{\bm C} $ only differs from $ {\bm C} $ in the  first row: instead of $(2,-1, 0, \cdots )$, we now have $(1,1, 0,\cdots)$.
 
\medskip
\noindent
We are now ready to demonstrate the minimality of the Lee-Yang vacuum character. For ease of discussion, we denote the exponent of $q$ in the first factor of \eqref{chivacn} as 
 \be
   Q_{k,h}(m_1, \cdots, m_k; n_1, \cdots n_h)=\bm m \bm C \bm m^T+\bm n_L\bm  B \bm n_R^T~,
 \ee
and in the first factor of  \eqref{chivacnii} as
 \be
 \tilde Q_{k,h}(m_1, \cdots, m_k; n_1, \cdots n_h)=\bm m \tilde{\bm C} \bm m^T+\bm n_L\bm  B \bm n_R^T+m_1~.
 \ee
 It is easy to see that  we have the relation
 \be\label{rel1}
    Q_{k,h}(m_1, \cdots, m_{k-1}, m_k=0; n_1, \cdots n_h)
  =   Q_{k-1,h}(m_1, \cdots, m_{k-1}; n_1, \cdots n_h)~.
 \ee
As a result, for $h,k>0$, \eqref{rel1} enables us to recursively reduce $[h+1,k+2]$ to $[h+1,2]$.  If $h=0$, we can reduce $[ 1,k+2]$ to $[ 1,3]$. Of course, we can also reduce further to $[1,2]$, but this is a trivial theory with zero central charge (technically speaking, it is only trivial once we quotient the Verma module by the null vectors). Note that we are not concerned with the $[1,k+2]$ theories, since these are unitary 2D theories and lead to non-unitary 4D theories (with $c<0$).\footnote{Nevertheless, we have   numerically checked to very high order---but have not tried to prove---that Lee-Yang has a smaller vacuum character than that of the CFT labelled by $[ 1,3]$ (i.e., the Ising model). If this is true, then the  Lee-Yang CFT also has smaller vacuum character than those of all unitary minimal models (i.e., those labeled by $[1, p]=(p,p+1)$).}
 
 Furthermore, using \eqref{chiNn}, one can show that
  \be
    Q_{0,h}( n_1, \cdots,n_{h-1}, n_h=0)
  =   Q_{0,h-1}( n_1, \cdots, n_{h-1})~.
 \ee
 Note that we have $m_1=0$ here.  So $[h+1,2]$  can further reduce to $[ 2,2]$, which is  the Lee-Yang theory we want to constrain (note again that $[1,2]$ is trivial). As a result, we have\footnote{One can explicitly see this by setting $m_i=0$ for all $i$ and $n_i=0$ for $i\neq 1$ in \eqref{chivacn}. The same discussion also applies to \eqref{minimality2}.}
\be
\chi_{1,1}^{[h+1,k+2]}> \chi^{[2, 2]}_{1,1}~, \qquad h,k>0~.
\ee
 
 Similarly, we also have
  \be\label{rel2}
  \tilde Q_{k,h}(m_1, \cdots, m_{k-1}, m_k=0; n_1, \cdots n_h)
  = \tilde Q_{k-1,h}(m_1, \cdots, m_{k-1}; n_1, \cdots n_h)
 \ee
 Therefore, if $h>0$, \eqref{rel2} enables us to reduce $[h+1, 1, k+1]$ to $[h+1, 1, 1]$, which is just $[h+1,2]$.  But this case has already been considered, and so
 \be\label{minimality2}
 \chi_{1,1}^{[h+1, 1, k+1]}>\chi_{1,1}^{[h+1,2]}>\chi_{1,1}^{[2,2]}~.
 \ee
 
Finally, if $h=0$, we can reduce $[ 1, 1, k+1]$ to $[1,1,2]$, which is just the $(3,5)$ minimal model. This theory has vacuum character
   \be
 \chi^{[1,1,2]}_{1,1}=\sum_{n=0}^\infty \frac{q^{n^2+n}}{(q)_{2n}}
 =\sum_{n=0}^\infty \frac{q^{n^2+n}}{(q)_{ n}} \prod_{j=n+1}^{2n}\frac{1}{\Big( 1-q^j\Big)}
 =\sum_{n=0}^\infty \frac{q^{n^2+n}}{(q)_{ n}}  (1+\text{positive series})~.
  \ee
Comparing with the expression for Lee-Yang,  
 \be
 \chi^{[2,2]}=\sum_{n=0}^\infty \frac{q^{n^2+n}}{(q)_n}~,
 \ee
 it is straightforward to conclude that 
 \be
   \chi_{1,1}^{[1,1,2]}>\chi_{1,1}^{[2,2]}~.
 \ee
 
 Therefore, we see that the Lee-Yang theory has the smallest vacuum character among all non-unitary minimal models. As we have discussed in the introduction to this section, this result implies that the MAD SCFT has the smallest Schur index of any unitary 4D $\CN=2$ SCFT with purely bosonic Schur operators (i.e., Claim 4). Moreover, this theory has the smallest number of Schur operators of any unitary 4D $\CN=2$ SCFT for every possible $h$ (i.e., Corollary 5). 

Combined with the minimality of the 4D central charge among unitary interacting SCFTs and our results on the equivalence of the spectral spin thresholds of the $\hat\CC_{R(j,j)}$ multiplets of the MAD theory with those of super Maxwell theory, we see that the MAD SCFT occupies a very special point in theory space.

\newsec{Conclusions}
In this paper, we focused on certain typically overlooked Schur multiplets, the $\hat\CC_{R(j,j)}$ multiplets with $R>0$. We showed that spin thresholds in this part of the Schur sector encode RG flows between free vector(s), the MAD SCFT, and its higher rank $(A_1, A_{2k})$ generalizations. Such information is more conventionally encoded in the Seiberg-Witten geometry. Typically, this geometry is associated with the ring structure of the $\CN=2$ chiral primaries. It would be interesting to understand if there is an intrinsically geometrical interpretation of the $\hat\CC_{R(j,j)}$ multiplets as well.

We also saw that the MAD theory has the smallest number of Schur operators for every $h$ among all local unitary 4D SCFTs. It would be of interest to try to generalize this notion beyond the Schur sector.\footnote{This question is subtle because non-Schur operators do not typically have half-integer quantized $U(1)_r$.}

Our paper suggests several additional questions:

\begin{itemize}
\item Are the MAD SCFT's spin thresholds \eqref{MADgaps} maximal in the space of unitary interacting 4D $\CN=2$ SCFTs?
\item More generally, do the spin threshold equalities \eqref{UVIRgenkForm} carve out the $(A_1, A_{2k})$ theories among all unitary interacting theories? If so, can we relate the corresponding infinite set of vanishing OPE coefficients in the UV and the IR? Does this relation allow us to say something about the $(A_1, A_{2k})$ theories beyond the Schur sector?\footnote{See also \cite{Xie:2021omd,Song:2021dhu} for other ideas in this direction.} If the spin threshold equalities do not carve out the $(A_1, A_{2k})$ SCFTs, what are the other theories satisfying similar equalities?
\item Can we prove the RG flow map in \eqref{RGmapMAD} and the more heuristic one in \eqref{RGmap2k}?
\item Can we prove that, modulo RG flows with irrelevant IR gauging, the equalities in \eqref{UVIRgenkForm} generalize to inequalities with thresholds in the UV that are no larger than the corresponding thresholds in the IR (here we again have in mind flows onto pure Coulomb branches)?
\item Can we use the Schur spectrum to effectively read off the most general Coulomb branch RG flows (already, for the higher $(A_{N-1},A_{K-1})$ theories with ${\rm gcd}(N,K)=1$  and $N, K>2$ we expect the situation to be more complicated)? What does the relation with the more standard Seiberg-Witten results tell us about the Coulomb branch and the Schur sector?
\item Are the large spin thresholds we find in the MAD theory indicative of some interesting large $SU(2)_R$-weight effective theory along the lines of the $U(1)_r$ analysis in \cite{Hellerman:2021yqz,Hellerman:2021duh}?
\item We saw that the MAD theory has the smallest Schur index for theories with only bosonic Schur generators. It would be interesting to understand if the same holds for the Macdonald index. This statement looks more non-trivial to prove at the level of the chiral algebra due to the \emph{non}-conservation of $SU(2)_R $ weight  in the OPE.\footnote{Nevertheless, it is straightforward to show that the MAD theory has smaller Macdonald index than the rest of the ($A_1, A_{2k}$) theories using \eqref{A1A2k}. We  have also checked numerically that the MAD theory has smaller Macdonald index than many other theories, including many of the ($A_1,D_{2k+1}$) SCFTs whose chiral algebras are given by certain affine Kac-Moody algebras.}
\end{itemize}

\ack{We are grateful to S.~Giacomelli, I.~Runkel, and S.~Wood for helpful correspondence and discussions. M.~B.'s work was supported by the Royal Society under the grant, \lq\lq New Aspects of Conformal and Topological Field Theories Across Dimensions." M.~B.'s and H.~J.'s work was supported by the Royal Society under the grant, “Relations, Transformations, and Emergence in Quantum Field Theory” and the STFC under the grant, “String Theory, Gauge Theory and Duality.”  T.~N.'s work was supported by JSPS KAKENHI Grant Numbers JP18K13547 and JP21H04993.}

\newpage
\begin{appendices}
\section{Generating a $\hat{\mathcal{C}}_{1(0,0)}$ multiplet from the $\hat{\mathcal{B}}_1\times\hat{\mathcal{B}}_1$ OPE} \label{appA}
In this appendix, we elaborate on the discussion around \eqref{B1B1} and explain how an adjoint-valued $\hat\CC_{1(0,0)}$ multiplet can arise in a theory with a locally realized simple flavor symmetry factor (as we explained in the main text, this $\hat\CC_{1(0,0)}$ multiplet need not be new; indeed, it can be involved in a null relation with the $\hat\CC_{1(0,0)}$ multiplet arising in the OPE in \eqref{C0B1}).

The basic idea is to re-examine the expression for the superconformal primaries of the four-point function of current multiplet primaries studied in \cite{Beem:2013sza} 
\begin{equation}
f^{ABCD}(z)\equiv z_{12}^2z_{34}^2\langle J^A(z_1)J^B(z_2)J^C(z_3)J^D(z_4)\rangle=\sum_{{\bf\mathcal{R}}\in\otimes^2{\rm adj}}P_{\mathcal{R}}^{ABCD}f_{\mathcal{R}}(z)
\end{equation}
where $\mathcal{P}^{ABCD}_{\bf\mathcal{R}}$ is a projector onto the ${\bf\mathcal{R}}$ irreducible representation. In particular, we will focus on the adjoint channel as opposed to the singlet.

To that end, the relevant projector is
\begin{align}
(P_\text{adj})_{AB;\,CD} \equiv
 \frac{1}{C_\text{adj}}(T_E)_{AB}(T_E)_{CD}~,
\label{eq:projector}
\end{align}
where $C_\text{adj}$ is the quadratic Casimir and $(T_E)_{AB}$ is the
representation matrix of the generator, both for the adjoint
representation \cite{Cvitanovic}. In terms of the structure constant
$f_{ABC}$, \eqref{eq:projector} can be rewritten as
\begin{align}
 (P_\text{adj})_{AB;CD} = -\frac{1}{C_\text{adj}}f_{EAB}f_{ECD}~.
\end{align}
Since $(P_\text{adj})^{AB}{}_{AB} = -\,\text{dim}\,G$, we find
\begin{align}
  f_\text{adj}(z) &= -\frac{1}{\text{dim}\,G}(P_\text{adj})_{AB;\,CD} f^{ABCD}(z)~.
\end{align}
Using the expression for $f^{ABCD}(z)$ shown in Eq.~(4.9) of \cite{Beem:2013sza},
we find\footnote{Here we used $f_{EAB}f_{EAB} = (\text{dim}\,G)C_\text{adj}$ and $f_{I_1I_2A}f_{I_2I_3B}f_{I_3I_1C} = \frac{1}{2}C_\text{adj}f_{ABC}$}
\begin{align}
 f_\text{adj}(z) =
 \frac{z-2}{z-1}\left(\frac{z^3}{z-1}-\frac{z}{2k_{2d}}C_{\text{adj}}\right)~,
\label{eq:fadj}
\end{align}
where $k_{2d}$ is the level of the 2D affine current corresponding to
the 4D flavor current multiplet.
In the conventions of \cite{Beem:2013sza}, we take the normalization that the long
root has length $\sqrt{2}$, which implies that $C_\text{adj}$ is related
to the dual coxeter number $h^\vee$ by $C_\text{adj} =
2h^\vee$. Moreover, $k_{2d}$ is related to the 4D flavor central charge
$k$ by \eqref{eq:centralcharges}. Therefore \eqref{eq:fadj} can be
rewritten as
\begin{align}
 f_\text{adj}(z) = \frac{z-2}{z-1}\left(\frac{z^3}{z-1}
 + \frac{2zh^\vee}{k}\right)~.
\end{align}
This function of $z$ can be decomposed as
\begin{align}
 f_\text{adj}(z) = \sum_{i=1}^\infty a_i g_i(z)~,\qquad g_i(z)\equiv
 \left(-\frac{z}{2}\right)^{i-1}z \,{}_2F_1(i,i;2i;z)~.
\end{align}
The first few non-vanishing coefficients are shown below:
\begin{align}
 a_1 &= \frac{4h^\vee}{k}~,
\\
a_3 &= \frac{8(h^\vee-3k)}{3k}~,
\\
a_5 &= \frac{32(h^\vee-10k)}{35k}~,
\\
a_7 &= \frac{64(h^\vee-21k)}{231k}~.
\end{align}

As shown in Eq.~(4.11) of \cite{Beem:2013sza}, $g_1(z)$ and $g_2(z)$ are the
contributions of $\hat{\mathcal{B}}_1$ and $\hat{\mathcal{B}}_2$,
respectively. Similarly, $g_{i\geq 3}(z)$ is the contribution of
$\hat{\mathcal{C}}_{1(\frac{i-3}{2},\frac{i-3}{2})}$ (where we assume the
absence of
higher spin multiplets $\hat{\mathcal{C}}_{0(j,j)}$ for $j>0$).
Therefore, the OPE coefficient for $\hat{\mathcal{C}}_{1(0,0)}$ is
\begin{align}
 a_3 = \frac{8(h^\vee-3k)}{3k}~,
\label{eq:coef}
\end{align}
up to a non-vanishing prefactor.
Note that, for a flavor symmetry group associated with a simple Lie algebra, Table
3 of \cite{Beem:2013sza} implies that unitarity puts a lower bound on
$k$ depending on the choice of the Lie algebra.\footnote{In
\cite{Beem:2013sza}, our $k$ is denoted by $k_{4d}$.} We see that
all these bounds force $k$ to satisfy
\begin{align}
 k>\frac{h^\vee}{3}~.
\end{align}
This implies that \eqref{eq:coef} is always non-vanishing, and therefore
$\hat{\mathcal{C}}_{1(0,0)}$ must be present in theories with a
non-trivial simple flavor symmetry factor.

\end{appendices}

\newpage
\bibliography{chetdocbib}

\begin{thebibliography}{10}
\ifx\href\asklfhas\newcommand{\href}[2]{#2}\fi
\ifx\arxivref\asklfhas\newcommand{\arxivref}[2]{\href{http://arxiv.org/abs/#1}{#2}}\fi
\ifx\doiref\asklfhas\newcommand{\doiref}[2]{\href{http://dx.doi.org/#1}{#2}}\fi
\parskip 0pt
\normalsize

\bibitem{Beem:2013sza}
C.~Beem, M.~Lemos, P.~Liendo, W.~Peelaers, L.~Rastelli \& B.~C. van~Rees,
\textit{``{Infinite Chiral Symmetry in Four Dimensions}''},
\doiref{10.1007/s00220-014-2272-x}{Commun.~Math.~Phys. \textbf{336}, 1359
  (2015)\ignorespaces}\ignorespaces,
\normalsize{\texttt{\arxivref{1312.5344}{arXiv:1312.5344
  \![hep-th]}}}\ignorespaces
\bibitem{Lemos:2014lua}
M.~Lemos \& W.~Peelaers,
\textit{``{Chiral Algebras for Trinion Theories}''},
\doiref{10.1007/JHEP02(2015)113}{JHEP \textbf{1502}, 113
  (2015)\ignorespaces}\ignorespaces,
\normalsize{\texttt{\arxivref{1411.3252}{arXiv:1411.3252
  \![hep-th]}}}\ignorespaces
\bibitem{Buican:2015ina}
M.~Buican \& T.~Nishinaka,
\textit{``{On the superconformal index of Argyres\textendash{}Douglas
  theories}''},
\doiref{10.1088/1751-8113/49/1/015401}{J.~Phys.~A \textbf{49}, 015401
  (2016)\ignorespaces}\ignorespaces,
\normalsize{\texttt{\arxivref{1505.05884}{arXiv:1505.05884
  \![hep-th]}}}\ignorespaces
\bibitem{Cordova:2015nma}
C.~Cordova \& S.-H. Shao,
\textit{``{Schur Indices, BPS Particles, and Argyres-Douglas Theories}''},
\doiref{10.1007/JHEP01(2016)040}{JHEP \textbf{1601}, 040
  (2016)\ignorespaces}\ignorespaces,
\normalsize{\texttt{\arxivref{1506.00265}{arXiv:1506.00265
  \![hep-th]}}}\ignorespaces
\bibitem{Buican:2016arp}
M.~Buican \& T.~Nishinaka,
\textit{``{Conformal Manifolds in Four Dimensions and Chiral Algebras}''},
\doiref{10.1088/1751-8113/49/46/465401}{J.~Phys.~A \textbf{49}, 465401
  (2016)\ignorespaces}\ignorespaces,
\normalsize{\texttt{\arxivref{1603.00887}{arXiv:1603.00887
  \![hep-th]}}}\ignorespaces
\bibitem{Xie:2016evu}
D.~Xie, W.~Yan \& S.-T. Yau,
\textit{``{Chiral algebra of the Argyres-Douglas theory from M5 branes}''},
\doiref{10.1103/PhysRevD.103.065003}{Phys.~Rev.~D \textbf{103}, 065003
  (2021)\ignorespaces}\ignorespaces,
\normalsize{\texttt{\arxivref{1604.02155}{arXiv:1604.02155
  \![hep-th]}}}\ignorespaces
\bibitem{Creutzig:2017qyf}
T.~Creutzig,
\textit{``{W-algebras for Argyres-Douglas theories}''},
\normalsize{\texttt{\arxivref{1701.05926}{arXiv:1701.05926
  \![hep-th]}}}\ignorespaces
\bibitem{Song:2017oew}
J.~Song, D.~Xie \& W.~Yan,
\textit{``{Vertex operator algebras of Argyres-Douglas theories from
  M5-branes}''},
\doiref{10.1007/JHEP12(2017)123}{JHEP \textbf{1712}, 123
  (2017)\ignorespaces}\ignorespaces,
\normalsize{\texttt{\arxivref{1706.01607}{arXiv:1706.01607
  \![hep-th]}}}\ignorespaces
\bibitem{Buican:2017fiq}
M.~Buican, Z.~Laczko \& T.~Nishinaka,
\textit{``{$ \mathcal{N} $ = 2 S-duality revisited}''},
\doiref{10.1007/JHEP09(2017)087}{JHEP \textbf{1709}, 087
  (2017)\ignorespaces}\ignorespaces,
\normalsize{\texttt{\arxivref{1706.03797}{arXiv:1706.03797
  \![hep-th]}}}\ignorespaces
\bibitem{Buican:2017rya}
M.~Buican \& Z.~Laczko,
\textit{``{Nonunitary Lagrangians and unitary non-Lagrangian conformal field
  theories}''},
\doiref{10.1103/PhysRevLett.120.081601}{Phys.~Rev.~Lett. \textbf{120}, 081601
  (2018)\ignorespaces}\ignorespaces,
\normalsize{\texttt{\arxivref{1711.09949}{arXiv:1711.09949
  \![hep-th]}}}\ignorespaces
\bibitem{Creutzig:2018lbc}
T.~Creutzig,
\textit{``{Logarithmic W-algebras and Argyres-Douglas theories at higher
  rank}''},
\doiref{10.1007/JHEP11(2018)188}{JHEP \textbf{1811}, 188
  (2018)\ignorespaces}\ignorespaces,
\normalsize{\texttt{\arxivref{1809.01725}{arXiv:1809.01725
  \![hep-th]}}}\ignorespaces
\bibitem{Bonetti:2018fqz}
F.~Bonetti, C.~Meneghelli \& L.~Rastelli,
\textit{``{VOAs labelled by complex reflection groups and 4d SCFTs}''},
\doiref{10.1007/JHEP05(2019)155}{JHEP \textbf{1905}, 155
  (2019)\ignorespaces}\ignorespaces,
\normalsize{\texttt{\arxivref{1810.03612}{arXiv:1810.03612
  \![hep-th]}}}\ignorespaces
\bibitem{Arakawa:2018egx}
T.~Arakawa,
\textit{``{Chiral algebras of class $\mathcal{S}$ and Moore-Tachikawa
  symplectic varieties}''},
\normalsize{\texttt{\arxivref{1811.01577}{arXiv:1811.01577
  \![math.RT]}}}\ignorespaces
\bibitem{Xie:2019yds}
D.~Xie \& W.~Yan,
\textit{``{W algebras, cosets and VOAs for 4d $ \mathcal{N} $ = 2 SCFTs from M5
  branes}''},
\doiref{10.1007/JHEP04(2021)076}{JHEP \textbf{2104}, 076
  (2021)\ignorespaces}\ignorespaces,
\normalsize{\texttt{\arxivref{1902.02838}{arXiv:1902.02838
  \![hep-th]}}}\ignorespaces
\bibitem{Xie:2019zlb}
D.~Xie \& W.~Yan,
\textit{``{Schur sector of Argyres-Douglas theory and $W$-algebra}''},
\doiref{10.21468/SciPostPhys.10.3.080}{SciPost~Phys. \textbf{10}, 080
  (2021)\ignorespaces}\ignorespaces,
\normalsize{\texttt{\arxivref{1904.09094}{arXiv:1904.09094
  \![hep-th]}}}\ignorespaces
\bibitem{Beem:2019snk}
C.~Beem, C.~Meneghelli, W.~Peelaers \& L.~Rastelli,
\textit{``{VOAs and rank-two instanton SCFTs}''},
\doiref{10.1007/s00220-020-03746-9}{Commun.~Math.~Phys. \textbf{377}, 2553
  (2020)\ignorespaces}\ignorespaces,
\normalsize{\texttt{\arxivref{1907.08629}{arXiv:1907.08629
  \![hep-th]}}}\ignorespaces
\bibitem{Xie:2019vzr}
D.~Xie \& W.~Yan,
\textit{``{4d $\mathcal{N}=2$ SCFTs and lisse W-algebras}''},
\doiref{10.1007/JHEP04(2021)271}{JHEP \textbf{2104}, 271
  (2021)\ignorespaces}\ignorespaces,
\normalsize{\texttt{\arxivref{1910.02281}{arXiv:1910.02281
  \![hep-th]}}}\ignorespaces
\bibitem{Beem:2017ooy}
C.~Beem \& L.~Rastelli,
\textit{``{Vertex operator algebras, Higgs branches, and modular differential
  equations}''},
\doiref{10.1007/JHEP08(2018)114}{JHEP \textbf{1808}, 114
  (2018)\ignorespaces}\ignorespaces,
\normalsize{\texttt{\arxivref{1707.07679}{arXiv:1707.07679
  \![hep-th]}}}\ignorespaces
\bibitem{Beem:2019tfp}
C.~Beem, C.~Meneghelli \& L.~Rastelli,
\textit{``{Free Field Realizations from the Higgs Branch}''},
\doiref{10.1007/JHEP09(2019)058}{JHEP \textbf{1909}, 058
  (2019)\ignorespaces}\ignorespaces,
\normalsize{\texttt{\arxivref{1903.07624}{arXiv:1903.07624
  \![hep-th]}}}\ignorespaces
\bibitem{Pan:2021ulr}
Y.~Pan, Y.~Wang \& H.~Zheng,
\textit{``{Defects, modular differential equations, and free field realization
  of $\mathcal N$ = 4 VOAs}''},
\normalsize{\texttt{\arxivref{2104.12180}{arXiv:2104.12180
  \![hep-th]}}}\ignorespaces
\bibitem{Seiberg:1994aj}
N.~Seiberg \& E.~Witten,
\textit{``{Monopoles, duality and chiral symmetry breaking in N=2
  supersymmetric QCD}''},
\doiref{10.1016/0550-3213(94)90214-3}{Nucl.~Phys.~B \textbf{431}, 484
  (1994)\ignorespaces}\ignorespaces,
\normalsize{\texttt{\arxivref{hep-th/9408099}{hep-th/9408099}}}\ignorespaces
\bibitem{Argyres:2015ffa}
P.~Argyres, M.~Lotito, Y.~L\"u \& M.~Martone,
\textit{``{Geometric constraints on the space of $ \mathcal{N} $ = 2 SCFTs.
  Part I: physical constraints on relevant deformations}''},
\doiref{10.1007/JHEP02(2018)001}{JHEP \textbf{1802}, 001
  (2018)\ignorespaces}\ignorespaces,
\normalsize{\texttt{\arxivref{1505.04814}{arXiv:1505.04814
  \![hep-th]}}}\ignorespaces
\bibitem{Argyres:2015gha}
P.~C. Argyres, M.~Lotito, Y.~L\"u \& M.~Martone,
\textit{``{Geometric constraints on the space of $ \mathcal{N} $ = 2 SCFTs.
  Part II: construction of special K\"ahler geometries and RG flows}''},
\doiref{10.1007/JHEP02(2018)002}{JHEP \textbf{1802}, 002
  (2018)\ignorespaces}\ignorespaces,
\normalsize{\texttt{\arxivref{1601.00011}{arXiv:1601.00011
  \![hep-th]}}}\ignorespaces
\bibitem{Moore:2017cmm}
G.~W. Moore \& I.~Nidaiev,
\textit{``{The Partition Function Of Argyres-Douglas Theory On A
  Four-Manifold}''},
\normalsize{\texttt{\arxivref{1711.09257}{arXiv:1711.09257
  \![hep-th]}}}\ignorespaces
\bibitem{Martone:2020nsy}
M.~Martone,
\textit{``{Towards the classification of rank-r$ \mathcal{N} $ = 2 SCFTs. Part
  I. Twisted partition function and central charge formulae}''},
\doiref{10.1007/JHEP12(2020)021}{JHEP \textbf{2012}, 021
  (2020)\ignorespaces}\ignorespaces,
\normalsize{\texttt{\arxivref{2006.16255}{arXiv:2006.16255
  \![hep-th]}}}\ignorespaces
\bibitem{Argyres:2020wmq}
P.~C. Argyres \& M.~Martone,
\textit{``{Towards a classification of rank r$ \mathcal{N} $ = 2 SCFTs. Part
  II. Special Kahler stratification of the Coulomb branch}''},
\doiref{10.1007/JHEP12(2020)022}{JHEP \textbf{2012}, 022
  (2020)\ignorespaces}\ignorespaces,
\normalsize{\texttt{\arxivref{2007.00012}{arXiv:2007.00012
  \![hep-th]}}}\ignorespaces
\bibitem{Cecotti:2021ouq}
S.~Cecotti, M.~Del~Zotto, M.~Martone \& R.~Moscrop,
\textit{``{The Characteristic Dimension of Four-dimensional $\mathcal{N} = 2$
  SCFTs}''},
\normalsize{\texttt{\arxivref{2108.10884}{arXiv:2108.10884
  \![hep-th]}}}\ignorespaces
\bibitem{Buican:2015hsa}
M.~Buican \& T.~Nishinaka,
\textit{``{Argyres\textendash{}Douglas theories, S$^1$ reductions, and
  topological symmetries}''},
\doiref{10.1088/1751-8113/49/4/045401}{J.~Phys.~A \textbf{49}, 045401
  (2016)\ignorespaces}\ignorespaces,
\normalsize{\texttt{\arxivref{1505.06205}{arXiv:1505.06205
  \![hep-th]}}}\ignorespaces
\bibitem{Fredrickson:2017yka}
L.~Fredrickson, D.~Pei, W.~Yan \& K.~Ye,
\textit{``{Argyres-Douglas Theories, Chiral Algebras and Wild Hitchin
  Characters}''},
\doiref{10.1007/JHEP01(2018)150}{JHEP \textbf{1801}, 150
  (2018)\ignorespaces}\ignorespaces,
\normalsize{\texttt{\arxivref{1701.08782}{arXiv:1701.08782
  \![hep-th]}}}\ignorespaces
\bibitem{Dedushenko:2019mnd}
M.~Dedushenko \& Y.~Wang,
\textit{``{4d/2d $\rightarrow $ 3d/1d: A song of protected operator
  algebras}''},
\normalsize{\texttt{\arxivref{1912.01006}{arXiv:1912.01006
  \![hep-th]}}}\ignorespaces
\bibitem{Gaiotto:2010be}
D.~Gaiotto, G.~W. Moore \& A.~Neitzke,
\textit{``{Framed BPS States}''},
\doiref{10.4310/ATMP.2013.v17.n2.a1}{Adv.~Theor.~Math.~Phys. \textbf{17}, 241
  (2013)\ignorespaces}\ignorespaces,
\normalsize{\texttt{\arxivref{1006.0146}{arXiv:1006.0146
  \![hep-th]}}}\ignorespaces
\bibitem{Cecotti:2010fi}
S.~Cecotti, A.~Neitzke \& C.~Vafa,
\textit{``{R-Twisting and 4d/2d Correspondences}''},
\normalsize{\texttt{\arxivref{1006.3435}{arXiv:1006.3435
  \![hep-th]}}}\ignorespaces
\bibitem{Iqbal:2012xm}
A.~Iqbal \& C.~Vafa,
\textit{``{BPS Degeneracies and Superconformal Index in Diverse Dimensions}''},
\doiref{10.1103/PhysRevD.90.105031}{Phys.~Rev.~D \textbf{90}, 105031
  (2014)\ignorespaces}\ignorespaces,
\normalsize{\texttt{\arxivref{1210.3605}{arXiv:1210.3605
  \![hep-th]}}}\ignorespaces
\bibitem{Cecotti:2015lab}
S.~Cecotti, J.~Song, C.~Vafa \& W.~Yan,
\textit{``{Superconformal Index, BPS Monodromy and Chiral Algebras}''},
\doiref{10.1007/JHEP11(2017)013}{JHEP \textbf{1711}, 013
  (2017)\ignorespaces}\ignorespaces,
\normalsize{\texttt{\arxivref{1511.01516}{arXiv:1511.01516
  \![hep-th]}}}\ignorespaces
\bibitem{Dedushenko:2018bpp}
M.~Dedushenko, S.~Gukov, H.~Nakajima, D.~Pei \& K.~Ye,
\textit{``{3d TQFTs from Argyres\textendash{}Douglas theories}''},
\doiref{10.1088/1751-8121/abb481}{J.~Phys.~A \textbf{53}, 43LT01
  (2020)\ignorespaces}\ignorespaces,
\normalsize{\texttt{\arxivref{1809.04638}{arXiv:1809.04638
  \![hep-th]}}}\ignorespaces
\bibitem{Buican:2019huq}
M.~Buican \& Z.~Laczko,
\textit{``{Rationalizing CFTs and Anyonic Imprints on Higgs Branches}''},
\doiref{10.1007/JHEP03(2019)025}{JHEP \textbf{1903}, 025
  (2019)\ignorespaces}\ignorespaces,
\normalsize{\texttt{\arxivref{1901.07591}{arXiv:1901.07591
  \![hep-th]}}}\ignorespaces
\bibitem{Argyres:1995jj}
P.~C. Argyres \& M.~R. Douglas,
\textit{``{New phenomena in SU(3) supersymmetric gauge theory}''},
\doiref{10.1016/0550-3213(95)00281-V}{Nucl.~Phys.~B \textbf{448}, 93
  (1995)\ignorespaces}\ignorespaces,
\normalsize{\texttt{\arxivref{hep-th/9505062}{hep-th/9505062}}}\ignorespaces
\bibitem{Liendo:2015ofa}
P.~Liendo, I.~Ramirez \& J.~Seo,
\textit{``{Stress-tensor OPE in $ \mathcal{N}=2 $ superconformal theories}''},
\doiref{10.1007/JHEP02(2016)019}{JHEP \textbf{1602}, 019
  (2016)\ignorespaces}\ignorespaces,
\normalsize{\texttt{\arxivref{1509.00033}{arXiv:1509.00033
  \![hep-th]}}}\ignorespaces
\bibitem{Castro-Alvaredo:2017udm}
O.~A. Castro-Alvaredo, B.~Doyon \& F.~Ravanini,
\textit{``{Irreversibility of the renormalization group flow in non-unitary
  quantum field theory}''},
\doiref{10.1088/1751-8121/aa8a10}{J.~Phys.~A \textbf{50}, 424002
  (2017)\ignorespaces}\ignorespaces,
\normalsize{\texttt{\arxivref{1706.01871}{arXiv:1706.01871
  \![hep-th]}}}\ignorespaces
\bibitem{Closset:2020scj}
C.~Closset, S.~Schafer-Nameki \& Y.-N. Wang,
\textit{``{Coulomb and Higgs Branches from Canonical Singularities: Part 0}''},
\doiref{10.1007/JHEP02(2021)003}{JHEP \textbf{2102}, 003
  (2021)\ignorespaces}\ignorespaces,
\normalsize{\texttt{\arxivref{2007.15600}{arXiv:2007.15600
  \![hep-th]}}}\ignorespaces
\bibitem{DelZotto:2020esg}
M.~Del~Zotto, I.~n. Garc\'\i{}a~Etxebarria \& S.~S. Hosseini,
\textit{``{Higher form symmetries of Argyres-Douglas theories}''},
\doiref{10.1007/JHEP10(2020)056}{JHEP \textbf{2010}, 056
  (2020)\ignorespaces}\ignorespaces,
\normalsize{\texttt{\arxivref{2007.15603}{arXiv:2007.15603
  \![hep-th]}}}\ignorespaces
\bibitem{Closset:2020afy}
C.~Closset, S.~Giacomelli, S.~Sch\"afer-Nameki \& Y.-N. Wang,
\textit{``{5d and 4d SCFTs: Canonical Singularities, Trinions and
  S-Dualities}''},
\normalsize{\texttt{\arxivref{2012.12827}{arXiv:2012.12827
  \![hep-th]}}}\ignorespaces
\bibitem{Closset:2021lwy}
C.~Closset, S.~Sch\"afer-Nameki \& Y.-N. Wang,
\textit{``{Coulomb and Higgs Branches from Canonical Singularities, Part 1:
  Hypersurfaces with Smooth Calabi-Yau Resolutions}''},
\normalsize{\texttt{\arxivref{2111.13564}{arXiv:2111.13564
  \![hep-th]}}}\ignorespaces
\bibitem{Buican:2021xhs}
M.~Buican \& H.~Jiang,
\textit{``{1-Form Symmetry, Isolated N=2 SCFTs, and Calabi-Yau Threefolds}''},
\normalsize{\texttt{\arxivref{2106.09807}{arXiv:2106.09807
  \![hep-th]}}}\ignorespaces
\bibitem{Song:2016yfd}
J.~Song,
\textit{``{Macdonald Index and Chiral Algebra}''},
\doiref{10.1007/JHEP08(2017)044}{JHEP \textbf{1708}, 044
  (2017)\ignorespaces}\ignorespaces,
\normalsize{\texttt{\arxivref{1612.08956}{arXiv:1612.08956
  \![hep-th]}}}\ignorespaces
\bibitem{Argyres:2012fu}
P.~C. Argyres, K.~Maruyoshi \& Y.~Tachikawa,
\textit{``{Quantum Higgs branches of isolated N=2 superconformal field
  theories}''},
\doiref{10.1007/JHEP10(2012)054}{JHEP \textbf{1210}, 054
  (2012)\ignorespaces}\ignorespaces,
\normalsize{\texttt{\arxivref{1206.4700}{arXiv:1206.4700
  \![hep-th]}}}\ignorespaces
\bibitem{Gadde:2011uv}
A.~Gadde, L.~Rastelli, S.~S. Razamat \& W.~Yan,
\textit{``{Gauge Theories and Macdonald Polynomials}''},
\doiref{10.1007/s00220-012-1607-8}{Commun.~Math.~Phys. \textbf{319}, 147
  (2013)\ignorespaces}\ignorespaces,
\normalsize{\texttt{\arxivref{1110.3740}{arXiv:1110.3740
  \![hep-th]}}}\ignorespaces
\bibitem{Dolan:2002zh}
F.~A. Dolan \& H.~Osborn,
\textit{``{On short and semi-short representations for four-dimensional
  superconformal symmetry}''},
\doiref{10.1016/S0003-4916(03)00074-5}{Annals~Phys. \textbf{307}, 41
  (2003)\ignorespaces}\ignorespaces,
\normalsize{\texttt{\arxivref{hep-th/0209056}{hep-th/0209056}}}\ignorespaces
\bibitem{Dobrev:1985qv}
V.~K. Dobrev \& V.~B. Petkova,
\textit{``{All Positive Energy Unitary Irreducible Representations of Extended
  Conformal Supersymmetry}''},
\doiref{10.1016/0370-2693(85)91073-1}{Phys.~Lett.~B \textbf{162}, 127
  (1985)\ignorespaces}\ignorespaces
\bibitem{Cordova:2016emh}
C.~Cordova, T.~T. Dumitrescu \& K.~Intriligator,
\textit{``{Multiplets of Superconformal Symmetry in Diverse Dimensions}''},
\doiref{10.1007/JHEP03(2019)163}{JHEP \textbf{1903}, 163
  (2019)\ignorespaces}\ignorespaces,
\normalsize{\texttt{\arxivref{1612.00809}{arXiv:1612.00809
  \![hep-th]}}}\ignorespaces
\bibitem{Maldacena:2011jn}
J.~Maldacena \& A.~Zhiboedov,
\textit{``{Constraining Conformal Field Theories with A Higher Spin
  Symmetry}''},
\doiref{10.1088/1751-8113/46/21/214011}{J.~Phys.~A \textbf{46}, 214011
  (2013)\ignorespaces}\ignorespaces,
\normalsize{\texttt{\arxivref{1112.1016}{arXiv:1112.1016
  \![hep-th]}}}\ignorespaces
\bibitem{Alba:2015upa}
V.~Alba \& K.~Diab,
\textit{``{Constraining conformal field theories with a higher spin symmetry in
  $d > 3$ dimensions}''},
\doiref{10.1007/JHEP03(2016)044}{JHEP \textbf{1603}, 044
  (2016)\ignorespaces}\ignorespaces,
\normalsize{\texttt{\arxivref{1510.02535}{arXiv:1510.02535
  \![hep-th]}}}\ignorespaces
\bibitem{Maruyoshi:2016tqk}
K.~Maruyoshi \& J.~Song,
\textit{``{Enhancement of Supersymmetry via Renormalization Group Flow and the
  Superconformal Index}''},
\doiref{10.1103/PhysRevLett.118.151602}{Phys.~Rev.~Lett. \textbf{118}, 151602
  (2017)\ignorespaces}\ignorespaces,
\normalsize{\texttt{\arxivref{1606.05632}{arXiv:1606.05632
  \![hep-th]}}}\ignorespaces
\bibitem{Agarwal:2018zqi}
P.~Agarwal, S.~Lee \& J.~Song,
\textit{``{Vanishing OPE Coefficients in 4d $N=2$ SCFTs}''},
\doiref{10.1007/JHEP06(2019)102}{JHEP \textbf{1906}, 102
  (2019)\ignorespaces}\ignorespaces,
\normalsize{\texttt{\arxivref{1812.04743}{arXiv:1812.04743
  \![hep-th]}}}\ignorespaces
\bibitem{Song:2015wta}
J.~Song,
\textit{``{Superconformal indices of generalized Argyres-Douglas theories from
  2d TQFT}''},
\doiref{10.1007/JHEP02(2016)045}{JHEP \textbf{1602}, 045
  (2016)\ignorespaces}\ignorespaces,
\normalsize{\texttt{\arxivref{1509.06730}{arXiv:1509.06730
  \![hep-th]}}}\ignorespaces
\bibitem{Foda:2019guo}
O.~Foda \& R.-D. Zhu,
\textit{``{Closed form fermionic expressions for the Macdonald index}''},
\doiref{10.1007/JHEP06(2020)157}{JHEP \textbf{2006}, 157
  (2020)\ignorespaces}\ignorespaces,
\normalsize{\texttt{\arxivref{1912.01896}{arXiv:1912.01896
  \![hep-th]}}}\ignorespaces
\bibitem{Argyres:1995xn}
P.~C. Argyres, M.~R. Plesser, N.~Seiberg \& E.~Witten,
\textit{``{New N=2 superconformal field theories in four-dimensions}''},
\doiref{10.1016/0550-3213(95)00671-0}{Nucl.~Phys.~B \textbf{461}, 71
  (1996)\ignorespaces}\ignorespaces,
\normalsize{\texttt{\arxivref{hep-th/9511154}{hep-th/9511154}}}\ignorespaces
\bibitem{Benvenuti:2017lle}
S.~Benvenuti \& S.~Giacomelli,
\textit{``{Supersymmetric gauge theories with decoupled operators and chiral
  ring stability}''},
\doiref{10.1103/PhysRevLett.119.251601}{Phys.~Rev.~Lett. \textbf{119}, 251601
  (2017)\ignorespaces}\ignorespaces,
\normalsize{\texttt{\arxivref{1706.02225}{arXiv:1706.02225
  \![hep-th]}}}\ignorespaces
\bibitem{Benvenuti:2017kud}
S.~Benvenuti \& S.~Giacomelli,
\textit{``{Abelianization and sequential confinement in $2+1$ dimensions}''},
\doiref{10.1007/JHEP10(2017)173}{JHEP \textbf{1710}, 173
  (2017)\ignorespaces}\ignorespaces,
\normalsize{\texttt{\arxivref{1706.04949}{arXiv:1706.04949
  \![hep-th]}}}\ignorespaces
\bibitem{Xie:2013jc}
D.~Xie \& P.~Zhao,
\textit{``{Central charges and RG flow of strongly-coupled N=2 theory}''},
\doiref{10.1007/JHEP03(2013)006}{JHEP \textbf{1303}, 006
  (2013)\ignorespaces}\ignorespaces,
\normalsize{\texttt{\arxivref{1301.0210}{arXiv:1301.0210
  \![hep-th]}}}\ignorespaces
\bibitem{Kiyoshige:2018wol}
K.~Kiyoshige \& T.~Nishinaka,
\textit{``{OPE Selection Rules for Schur Multiplets in 4D $\mathcal{N}=2$
  Superconformal Field Theories}''},
\doiref{10.1007/JHEP04(2019)060}{JHEP \textbf{1904}, 060
  (2019)\ignorespaces}\ignorespaces,
\normalsize{\texttt{\arxivref{1812.06394}{arXiv:1812.06394
  \![hep-th]}}}\ignorespaces
\bibitem{Nirschl:2004pa}
M.~Nirschl \& H.~Osborn,
\textit{``{Superconformal Ward identities and their solution}''},
\doiref{10.1016/j.nuclphysb.2005.01.013}{Nucl.~Phys.~B \textbf{711}, 409
  (2005)\ignorespaces}\ignorespaces,
\normalsize{\texttt{\arxivref{hep-th/0407060}{hep-th/0407060}}}\ignorespaces
\bibitem{DiFrancesco:1997nk}
P.~Di~Francesco, P.~Mathieu \& D.~Senechal,
\textit{``{Conformal Field Theory}''},
Springer-Verlag (1997)\ignorespaces,
New York
\bibitem{Buican:2020moo}
M.~Buican \& T.~Nishinaka,
\textit{``{$\mathcal{N}=4$ SYM, Argyres-Douglas Theories, and an Exact Graded
  Vector Space Isomorphism}''},
\normalsize{\texttt{\arxivref{2012.13209}{arXiv:2012.13209
  \![hep-th]}}}\ignorespaces
\bibitem{Kang:2021lic}
M.~J. Kang, C.~Lawrie \& J.~Song,
\textit{``{Infinitely many 4D N=2 SCFTs with a=c and beyond}''},
\doiref{10.1103/PhysRevD.104.105005}{Phys.~Rev.~D \textbf{104}, 105005
  (2021)\ignorespaces}\ignorespaces,
\normalsize{\texttt{\arxivref{2106.12579}{arXiv:2106.12579
  \![hep-th]}}}\ignorespaces
\bibitem{Welsh:2002jq}
T.~A. Welsh,
\textit{``{Fermionic expressions for minimal model Virasoro characters}''},
\normalsize{\texttt{\arxivref{math/0212154}{math/0212154}}}\ignorespaces
\bibitem{Gaberdiel:1996kx}
M.~R. Gaberdiel \& H.~G. Kausch,
\textit{``{Indecomposable fusion products}''},
\doiref{10.1016/0550-3213(96)00364-1}{Nucl.~Phys.~B \textbf{477}, 293
  (1996)\ignorespaces}\ignorespaces,
\normalsize{\texttt{\arxivref{hep-th/9604026}{hep-th/9604026}}}\ignorespaces
\bibitem{Melzer:1993zk}
E.~Melzer,
\textit{``{Fermionic character sums and the corner transfer matrix}''},
\doiref{10.1142/S0217751X94000510}{Int.~J.~Mod.~Phys.~A \textbf{9}, 1115
  (1994)\ignorespaces}\ignorespaces,
\normalsize{\texttt{\arxivref{hep-th/9305114}{hep-th/9305114}}}\ignorespaces
\bibitem{Xie:2021omd}
D.~Xie \& W.~Yan,
\textit{``{A study of N =1 SCFT derived from N =2 SCFT: index and chiral
  ring}''},
\normalsize{\texttt{\arxivref{2109.04090}{arXiv:2109.04090
  \![hep-th]}}}\ignorespaces
\bibitem{Song:2021dhu}
J.~Song,
\textit{``{Vanishing short multiplets in rank one 4d/5d SCFTs}''},
\normalsize{\texttt{\arxivref{2109.05588}{arXiv:2109.05588
  \![hep-th]}}}\ignorespaces
\bibitem{Hellerman:2021yqz}
S.~Hellerman \& D.~Orlando,
\textit{``{Large R-charge EFT correlators in N=2 SQCD}''},
\normalsize{\texttt{\arxivref{2103.05642}{arXiv:2103.05642
  \![hep-th]}}}\ignorespaces
\bibitem{Hellerman:2021duh}
S.~Hellerman,
\textit{``{On the exponentially small corrections to ${\cal N} = 2$
  superconformal correlators at large R-charge}''},
\normalsize{\texttt{\arxivref{2103.09312}{arXiv:2103.09312
  \![hep-th]}}}\ignorespaces
\bibitem{Cvitanovic}
P.~Cvitanovic,
\textit{``{Group theory: Birdtracks, Lie’s and exceptional groups}''},
Princeton University Press (2008)\ignorespaces,
Princeton
\end{thebibliography}

\end{document}